\documentclass{article}

\PassOptionsToPackage{numbers, compress}{natbib}

\usepackage[preprint]{neurips_2026}

\usepackage[utf8]{inputenc} %
\usepackage[T1]{fontenc}    %
\usepackage{hyperref}       %
\usepackage{url}            %
\usepackage{booktabs}       %
\usepackage{amsfonts}       %
\usepackage{nicefrac}       %
\usepackage{microtype}      %
\usepackage{xcolor}         %
\usepackage{graphicx}
\usepackage{bm}
\usepackage{algorithm}
\usepackage{siunitx}
\usepackage{algpseudocode}
\usepackage{amsmath}
\usepackage{tabularx}
\usepackage{multirow}

\algnewcommand{\Input}{\item[\textbf{Input:}]}
\algnewcommand{\Hyperparameters}{\item[\textbf{Hyperparameters:}]}

\title{Curriculum Multiple Shooting for Robust Training of Neural and Universal Differential Equations}

\author{
  Sebastian ~Persson\thanks{Equal contribution.} \\
  Dynamics of Living Systems Laboratory\\
  The Francis Crick Institute\\
  London, 1 Midland Rd, United Kingdom \\
  \And
  Giacomo ~Fabrini$^{*}$ \\
  Institute for Chemical and Bioengineering\\
  Department of Chemical and Applied Biosciences\\
  ETH Zurich\\
  Zurich, Vladimir-Prelog-Weg 1-5/10, 8093, Switzerland\\
  \texttt{giacomo.fabrini@chem.ethz.ch} \\
  \And
  Branwen ~Snelling \\
  Dynamics of Living Systems Laboratory\\
  The Francis Crick Institute\\
  London, 1 Midland Rd, United Kingdom \\
  \texttt{branwen.snelling@crick.ac.uk} \\
  \And
  Fabian Fröhlich\thanks{Corresponding author.} \\
  Dynamics of Living Systems Laboratory\\
  The Francis Crick Institute\\
  London, 1 Midland Rd, United Kingdom \\
  \texttt{fabian.frohlich@crick.ac.uk} \\
}

\begin{document}

\maketitle

\begin{abstract}
    Neural ordinary differential equations (NODEs) and universal differential equations (UDEs) provide flexible and popular frameworks for learning interpretable dynamical systems from noisy time-series data. However, training these models remains challenging, and versatile methods that robustly handle sparse and noisy data as well as partially observed models are lacking. To address this, we introduce curriculum multiple shooting (CMS), a general-purpose training strategy for fitting ordinary differential equation (ODE) models to time-series data by integrating curriculum learning with multiple shooting. Across twelve benchmarks spanning simulated and real data, and covering NODEs, UDEs, and mechanistic ODEs, CMS accelerates and stabilises training convergence, outperforms state-of-the-art training strategies, and ranks among the best methods in generalisation. Finally, we discuss possible explanations for the strong performance of CMS in light of contemporary theories of what makes training on time-series challenging.
\end{abstract}

\section{Introduction}

Using mathematical models to predict the time-evolution of dynamical systems is a central challenge across many scientific and engineering disciplines. A classical approach is to iteratively construct mechanistic ordinary differential equation (ODE) models and evaluate them by fitting them to time-series data. As this process is labour-intensive, there is growing interest in using machine learning (ML) to learn interpretable dynamical models directly from data. In particular, neural ODEs (NODEs) combine the widely used ODE-based framework to model dynamical systems with the flexibility of neural networks~\citep{chen_neural_2018}. Universal differential equations (UDEs), also called grey-box models or hybrid NODEs~\citep{brucker_neural_2022, zhang_two_2022, zou_hybrid2_2024}, extend this idea by modelling known components mechanistically and unknown components with neural networks~\citep{rackauckas_universal_2021}. This makes UDEs attractive when prior knowledge is available and/or when data are limited. Accordingly, NODEs and UDEs are increasingly used across domains including biology~\citep{de_rooij_conditional_2025, steinacker_developing_2025, wu_data-driven_2025}, pharmacology/medicine~\citep{zou_hybrid2_2024, qian_integrating_2021, lu_neural-ode_2021, bram_low-dimensional_2024}, epidemiology~\citep{dandekar_machine_2020, ye_integrating_2025, murgai_abm-ude_2026}, and engineering applications such as automotive modelling~\citep{thummerer_neuralfmu_2022}.

Despite their conceptual appeal, NODEs and UDEs are difficult to train on time-series data~\citep{ko_homotopy-based_2023, philipps_current_2025}. One line of research addresses this challenge by imposing restrictions on what the ML model can learn~\citep{choromanski_ode_2020, finlay_how_2020, rodriguez_lyanet_2022, philipps_non-negative_2024, luo_fxts-net_2025}. Here, we instead focus on complementary training strategies given a fixed model structure. Existing strategies include methods from mechanistic ODE training, such as collocation and multiple shooting~\citep{roesch_collocation_2021, turan_multiple_2022}, as well as ML-inspired strategies like curriculum learning and homotopy-based training~\citep{ko_homotopy-based_2023, rackauckas_diffeqfluxjl_2019}. However, these strategies often trade off applicability against practicality. For example, collocation-based approaches require densely sampled data and full observation of the ODE model states because they rely on accurate data interpolation~\citep{liang_parameter_2008}. In many application domains~\citep{bram_low-dimensional_2024, ye_integrating_2025, hass_benchmark_2019}, however, time-series data are noisy, sparse, irregularly sampled, and the model is often only partially observed. Partial observability can further arise from the model architecture itself, as in augmented NODEs~\citep{dupont_augmented_2019}. Meanwhile, hyperparameter tuning can be non-trivial for more broadly applicable training strategies such as multiple shooting~\citep{turan_multiple_2022}. Ideally, a training strategy should be broadly applicable across these settings and robust to tuning, allowing practitioners to focus on modelling rather than training.

\textbf{Our contributions} are threefold. First, we introduce \emph{curriculum multiple shooting} (CMS), a general-purpose training strategy for fitting mechanistic ODE, NODE, and UDE models to noisy time-series data in both fully and partially observed settings. Second, we benchmark this method against state-of-the-art training strategies on four NODE/UDE benchmarks using simulated and real data, as well as on eight real-data mechanistic ODE benchmarks. These benchmarks span partial observability, multiscale and stiff dynamics, and noisy or irregularly sampled time-series data. Third, we analyse optimisation traces and loss landscapes to gain intuition for why CMS often improves training.

\section{Related work}
\label{sec:related_work}

\textbf{Training mechanistic ODEs.}
A large literature addresses fitting mechanistic ODE models to time-series data. In biology, empirical benchmarks have found that multistart parameter estimation with local second-order optimizers applied to a single-shooting objective (Eq.~\ref{eq:objective}) performs well in practice~\citep{hass_benchmark_2019, raue_lessons_2013}. Beyond standard single shooting, other training strategies include multiple shooting~\citep{bock_multiple_1984}; collocation methods, which approximate state derivatives from interpolated data and match them to the ODE model right-hand side~\citep{liang_parameter_2008}; and prediction-error~\citep{ljung_prediction_2002} and synchronization-based methods~\citep{abarbanel_dynamical_2009}, both of which use data interpolation to guide model simulations toward the data. All of these strategies have also been tested on NODEs and/or UDEs~\citep{dandekar_machine_2020, murgai_abm-ude_2026, ko_homotopy-based_2023, philipps_non-negative_2024, roesch_collocation_2021, turan_multiple_2022}.

\textbf{Improving performance for NODEs and UDEs.}
Several approaches have been proposed to improve performance. One is to constrain what the ML component can learn, either through soft constraints in the objective via regularization or auxiliary loss terms~\citep{zou_hybrid2_2024, de_rooij_conditional_2025, finlay_how_2020, rodriguez_lyanet_2022, luo_fxts-net_2025, yin_augmenting_2021, aliee_sparsity_2022}, or through hard constraints encoded directly in the model structure~\citep{choromanski_ode_2020, philipps_non-negative_2024, yin_augmenting_2021, greydanus_hamiltonian_2019}. State augmentation has likewise been proposed to make the underlying dynamics easier to simulate~\citep{dupont_augmented_2019}. Several of these methods aim to stabilize training by learning better-conditioned dynamics, which can help mitigate training instabilities such as exploding/vanishing gradients~\citep{choromanski_ode_2020, finlay_how_2020, rodriguez_lyanet_2022, luo_fxts-net_2025}. Another line of work focuses on improving training via gradient computation through the ODE solver~\citep{onken_discretize-optimize_2020, kim_stiff_2021, kidger_neural_2022, sapienza_differentiable_2025}. Roughly, the two main gradient schemes are discretize-then-optimize and optimize-then-discretize. While benchmarks remain limited, the choice of gradient scheme has been suggested to affect training performance in addition to runtime~\citep{onken_discretize-optimize_2020}. In App.~\ref{sec:julia_ude_benchmark}, we show that our results are robust across both schemes. Complementary to these approaches, we focus on the training strategy itself. Beyond aforementioned methods such as multiple shooting~\citep{dandekar_machine_2020}, NODE/UDE-targeted strategies have also been proposed. Curriculum learning, often referred to as "growing fits", has been used but remains under-benchmarked~\citep{rackauckas_diffeqfluxjl_2019, laudo_stable_2026}. Related sliding-window training~\citep{wu_data-driven_2025} and homotopy-based synchronization~\citep{ko_homotopy-based_2023} have also been applied.

\section{Problem setup and preliminaries}
\label{sec:problem_preliminaries}

This section introduces the model and data setting, and the two training strategies (curriculum learning and multiple shooting) forming the basis of our curriculum multiple shooting strategy.

\subsection{Training setup for ODE-based models}
\label{sec:ude_setup}

We consider training ordinary differential equations (ODEs) augmented with neural networks on noisy time-series data. These models, known as universal differential equations (UDEs), also called grey-box models or hybrid NODEs~\citep{rackauckas_universal_2021}, describe the time evolution of a state vector $\mathbf{u} \in \mathbb{R}^M$ as:

\begin{equation}
    \label{eq:ude_def}
    \frac{\mathrm{d}\mathbf{u}}{\mathrm{d}t}
    =
    \mathbf{f}\big(\mathbf{u}, t, \bm{\theta}_m, \mathbf{NN}(\mathbf{u}, t, \bm{\theta}_n)\big),
    \quad
    \mathbf{u}(t_0)=\mathbf{u}_{t_0}(\bm{\theta}_m),
\end{equation}

where $\mathbf{f}$ combines known mechanistic structure with unknown mechanistic parameters $\bm{\theta}_m$ and a neural network $\mathbf{NN}$ with unknown parameters $\bm{\theta}_n$. In the absence of a mechanistic component, i.e., when $\mathbf{f} = \mathbf{NN}(\mathbf{u}, t, \bm{\theta}_n)$, Eq.~\ref{eq:ude_def} reduces to a standard NODE~\citep{chen_neural_2018}.

Training Eq.~\ref{eq:ude_def} corresponds to estimating the unknown parameters $(\bm{\theta}_m,\bm{\theta}_n)$ by fitting the model to noisy time-series data. We assume data are available for $O$ observables, where for each observable $o = 1, \ldots, O$, measurements are taken at time points $\{t_{o,j}\}_{j=1}^{T_o}$ and the corresponding model output is given by the possibly nonlinear observation function $h_o\!\left(\mathbf{u}(t_{o,j}), \bm{\theta}_m\right)$. In many applications, such as biology and pharmacology~\citep{hass_benchmark_2019, gabrielsson_pharmacokinetic_2001}, models are only partially observed ($O < M$) and data are irregularly sampled. Both settings are considered in our experiments (Sec.~\ref{sec:experiments}).

Given this observation model, the parameters in Eq.~\ref{eq:ude_def} are estimated by minimizing a single-shooting objective function that quantifies the mismatch between model output and time-series data:

\begin{equation}
    \label{eq:objective}
    g
    =
    \sum_{o=1}^{O}\sum_{j=1}^{T_o}
    g_o\!\left(h_o\!\left(\mathbf{u}(t_{o,j}), \bm{\theta}_m\right), \tilde{m}_{o,t_{o,j}}\right),
\end{equation}

where $\tilde{m}_{o,t_{o,j}}$ denotes the noisy measurement of observable $o$ at time $t_{o,j}$. Common choices for $g_o$ include mean squared error (MSE) and likelihood-based objectives. When fitting UDEs, optimizing only Eq.~\ref{eq:objective} can allow the neural component to dominate the dynamics and "cannibalise" the mechanistic component~\citep{whipple_hybrid_2024}. One way to mitigate this is to add neural-network output regularization~\citep{philipps_non-negative_2024}:

\begin{equation}
    \label{eq:output_reg}
    g_{\mathrm{out}}
    =
    \lambda_O
    \left(
    \int_{t_0}^{T}
    \left\|
    \mathbf{NN}(\mathbf{u}(t), t, \bm{\theta}_n)
    \right\|_2
    \mathrm{d}t
    \right)^2,
\end{equation}

where $\lambda_O$ controls the regularization strength. When used, $g_{\mathrm{out}}$ is added to the objective in Eq.~\ref{eq:objective}.

In general, evaluating Eq.~\ref{eq:objective} requires numerically solving Eq.~\ref{eq:ude_def} with an ODE solver. A possible alternative is to reconstruct interpolants $\tilde{\mathbf{u}}(t)$ and $\mathrm{d}\tilde{\mathbf{u}}/\mathrm{d}t$ from the measurements, and then directly compare $\mathrm{d}\tilde{\mathbf{u}}/\mathrm{d}t$ to the right-hand side of Eq.~\ref{eq:ude_def}. We do not consider that setting here, since it requires the full state $\mathbf{u}$ to be observed, which is uncommon in many application domains like biology~\citep{hass_benchmark_2019}.

\subsection{Curriculum learning for ODE-based models}
\label{sec:curriculum_learning}

Curriculum learning (CL) is a training strategy in which a model is trained over $n_c$ stages on progressively harder tasks~\citep{bengio_curriculum_2009, soviany_curriculum_2022}. What constitutes a harder task depends on the application. Since training often becomes more difficult over longer time horizons~\citep{ko_homotopy-based_2023, ribeiro_smoothness_2020} for differential equations of the form in Eq.~\ref{eq:ude_def}, a natural curriculum is to train on progressively longer time intervals.

More formally, if measurements are available on $[t_0,T]$, a curriculum with $n_c$ stages is defined by the intervals $[t_0,t_1], \ldots, [t_0,t_i], \ldots, [t_0,t_{n_c}]$, where $t_{n_c}=T$. At stage $i$, the model is trained using only measurements in $[t_0,t_i]$.

\subsection{Multiple shooting for ODE-based models}
\label{sec:multiple_shooting}

Multiple shooting (MS) is a training strategy in which the time interval $[t_0,T]$ is divided into $n_w$ windows, and the model is fitted separately on each window~\citep{turan_multiple_2022, bock_multiple_1984}. As with curriculum learning, the motivation is that fitting multiple shorter trajectories is typically easier than fitting a long trajectory.

To define the MS training objective, we partition the time interval $[t_0, T]$ into $n_w$ windows with boundaries $t_0 < t_1 < \cdots < t_{n_w} = T$. We denote the resulting set of windows by $\mathcal{W}_{n_w} = \{w_i\}_{i=1}^{n_w}$, where $w_i := [t_{i-1}, t_i]$. For ease of notation, we drop the observable index $o$ from Eq.~\ref{eq:objective}, write $(t_j, \tilde{m}_j)$ for the resulting measurement time points and values, and let $\mathcal{M}_i := \{\, t_j : t_j \in w_i \,\}$ denote the measurement time points in $w_i$. As we assume accurate interpolants of the model state $\mathbf{u}$ are unavailable (Sec.~\ref{sec:ude_setup}), we introduce additional parameters to estimate, $\bm{\theta}_{u_0,i}$, for the initial state of each window. The multiple-shooting objective is computed by solving the model on each window:

\begin{equation}
\label{eq:ms_ode}
\frac{\mathrm{d}\mathbf{u}^{(i)}}{\mathrm{d}t}
=
\mathbf{f}\big(\mathbf{u}^{(i)}, t, \bm{\theta}_m, \mathbf{NN}(\mathbf{u}^{(i)}, t, \bm{\theta}_n)\big),
\qquad
\mathbf{u}^{(i)}(t_{i-1}) = \bm{\theta}_{u_0,i},
\qquad
t \in w_i,
\end{equation}

followed by summing the data-fitting loss over windows:

\begin{equation}
\label{eq:ms_objective}
g_{\mathrm{MS}}
=
\sum_{i=1}^{n_w}\sum_{t_j \in \mathcal{M}_i}
g\!\left(h\!\left(\mathbf{u}^{(i)}(t_j), \bm{\theta}_m\right), \tilde{m}_j\right).
\end{equation}

To enforce continuity between windows, a quadratic penalty is typically added~\citep{murgai_abm-ude_2026, rackauckas_diffeqfluxjl_2019}. While plain MS penalizes discrepancies at the window boundary, to allow longer overlapping windows we define the penalty on shared measurement time points between windows, i.e.\ $\mathcal{P}_i := \mathcal{M}_i \cap \mathcal{M}_{i+1}$:

\begin{equation}
    \label{eq:ms_penalty}
    g_{\mathrm{cont}}
    =
    \lambda_{\mathrm{cont}}
    \sum_{i=1}^{n_w-1}
    \sum_{t \in \mathcal{P}_i}
    \left\|
        \mathbf{u}^{(i)}(t) - \mathbf{u}^{(i+1)}(t)
    \right\|_2^2.
\end{equation}

Here, $\lambda_{\mathrm{cont}}$ controls the penalty strength. While other ways of handling continuity have been proposed~\citep{turan_multiple_2022, bock_multiple_1984, massaroli_differentiable_2021}, we use this common formulation~\citep{rackauckas_diffeqfluxjl_2019}, which our CMS strategy builds on.

\section{Curriculum multiple shooting}
\label{sec:cl_ms}

Our curriculum multiple shooting (CMS) strategy aims to combine the strengths of curriculum learning (CL; Sec.~\ref{sec:curriculum_learning}) and multiple shooting (MS; Sec.~\ref{sec:multiple_shooting}) while mitigating their main limitations. A general challenge in CL is to design curriculum stages that are representative of the final objective~\citep{bengio_curriculum_2009, soviany_curriculum_2022}. When the time-series data in early CL stages are not representative of the full trajectory, any overfitting incurred there must be corrected during later stages, potentially slowing training. In contrast, MS exposes the full trajectory from the start, but never directly trains on the final single-shooting objective, instead relying on the hard-to-tune continuity penalty ($\lambda_{\mathrm{cont}}$ Eq.~\ref{eq:ms_penalty}) to make the MS objective a good surrogate. CMS addresses this by starting from the MS objective and progressively merging shooting windows across curriculum stages until recovering single shooting. This yields easier initial optimization on short trajectories, retains a full-trajectory view throughout training, and gradually transitions to the final objective which reduces sensitivity to $\lambda_{\mathrm{cont}}$.

Specifically, CMS starts from a multiple-shooting objective with $n_w$ windows spanning $[t_0,T]$. At curriculum stage $s \in \{1,\ldots,n_w\}$, the window set is $\mathcal{W}^{(s)} = \{\, [t_{i-1},\, t_{i+s-1}] \,\}_{i=1}^{n_w-s+1}$, so adjacent windows are merged into longer overlapping windows. The number of windows thereby decreases by one per curriculum stage, and at the final stage $s=n_w$ only a single window $[t_0,T]$ remains, recovering the standard single-shooting objective (Eq.~\ref{eq:objective}). When the windows grow, the data-fitting term (Eq.~\ref{eq:ms_objective}) increasingly promotes agreement between neighboring windows in addition to the continuity penalty (Eq.~\ref{eq:ms_penalty}), reducing sensitivity to $\lambda_{\mathrm{cont}}$. Pseudocode is given in Alg.~\ref{alg:CMS}.

\begin{algorithm}[ht]
\caption{Curriculum multiple shooting (CMS)}
\label{alg:CMS}
\begin{algorithmic}[1]
\Input Time-series data $\mathcal{D}=\{(t_j,\tilde m_j)\}_{j=1}^{M}$, initial model parameters $(\bm{\theta}_m^{(0)},\bm{\theta}_n^{(0)})$
\Hyperparameters Initial number of windows $n_w$, continuity penalty $\lambda_{\mathrm{cont}}$, cumulative stage epoch budget $\{E_s\}_{s=1}^{n_w}$ with $E_1 < \cdots < E_{n_w}$, learning rate $\eta$, initial window values $\{\bm{\theta}_{u_0,i}^{(1)}\}_{i=1}^{n_w}$
\State Initialize $(\bm{\theta}_m,\bm{\theta}_n)\gets(\bm{\theta}_m^{(0)},\bm{\theta}_n^{(0)})$
\State Set initial window boundaries $t_0 < t_1 < \cdots < t_{n_w}=T$ and let $\mathcal{W}^{(1)}=\{[t_{i-1},t_i]\}_{i=1}^{n_w}$
\For{$s=1,\dots,n_w$}
    \State $k^{(s)} \gets |\mathcal{W}^{(s)}|$ \Comment{At $s=n_w$, the loss reduces to single-shooting as $k^{(n_w)}=1$ }
    \For{$\mathrm{epoch}=E_{s-1}+1,\dots,E_s$} \Comment{$E_0 := 0$}
        \State Solve Eq.~\ref{eq:ms_ode} on all $w_i^{(s)} \in \mathcal{W}^{(s)}$ with $(\bm{\theta}_m,\bm{\theta}_n,\bm{\theta}_{u_0,i}^{(s)})$ to obtain $\mathbf{u}^{(i,s)}(t)$
        \State Form the stage objective by evaluating Eqs.~\ref{eq:ms_objective}-~\ref{eq:ms_penalty} on $\mathcal{W}^{(s)}$: $g_{\mathrm{CMS}}^{(s)} \gets g_{\mathrm{MS}}^{(s)} + g_{\mathrm{cont}}^{(s)}$
        \State Update $(\bm{\theta}_m,\bm{\theta}_n,\{\bm{\theta}_{u_0,i}^{(s)}\}_{i=1}^{n_w^{(s)}})$ with one optimizer step using $\nabla g_{\mathrm{CMS}}^{(s)}$ and $\eta$
    \EndFor
    \State Form longer overlapping windows for the next stage: $\mathcal{W}^{(s+1)} = \left\{[\,t_{i-1},\, t_{i+s}\,]\right\}_{i=1}^{n_w-s}$
    \State Set $\bm{\theta}_{u_0,i}^{(s+1)} \gets \bm{\theta}_{u_0,i}^{(s)}$ for $i=1,\dots,k^{(s)}-1$
\EndFor
\State \Return $(\bm{\theta}_m,\bm{\theta}_n)$
\end{algorithmic}
\end{algorithm}

While Alg.~\ref{alg:CMS} is presented with a gradient-based optimizer (e.g. Adam), the method is compatible with most local optimizers; for example, in the ODE experiments (Sec.~\ref{sec:experiments}) we use a Newton trust-region method. We also found that, instead of penalizing all shared measurement points (Eq.~\ref{eq:ms_penalty}), penalizing only the first overlapping point as in plain multiple shooting also works (App.~\ref{sec:julia_ude_benchmark}).

\subsection{Hyperparameters}

CMS introduces five hyperparameters. Based on our experiments (Sec.~\ref{sec:experiments} and App.~\ref{app:experiment_details}), we found that three require tuning, whereas the remaining two can be fixed to reasonable default values. To facilitate applying CMS, we summarize practical values here.

The hyperparameters requiring tuning are the \textbf{learning rate} ($\eta$), the \textbf{number of initial multiple-shooting windows} ($n_w$), and the \textbf{continuity penalty} ($\lambda_{\mathrm{cont}}$). For $\eta$, values used for conventional single-shooting training (Eq.~\ref{eq:objective}) were sufficient, around $\num{1e-3}$. The number of initial multiple-shooting windows, equivalently the number of curriculum stages, controls the initial window length; values of $5$ or larger were sufficient. The continuity penalty controls the strength of agreement between adjacent windows; values in the range $0$--$10$ were sufficient.

The remaining hyperparameters, for which we found reasonable defaults, are the \textbf{cumulative stage epoch budget} ($\{E_s\}_{s=1}^{n_w}$) and the \textbf{window-state initialization} ($\bm{\theta}_{u_0,i}^{(1)}$). We found it sufficient to set $\{E_s\}_{s=1}^{n_w}$ such that the first third of the total epoch budget ($E_{n_w}/3$) was allocated equally across the first $n_w-1$ curriculum stages, with the remaining two thirds allocated to the final single-shooting stage. For $\bm{\theta}_{u_0,i}^{(1)}$, setting all initial window values to a small constant (e.g. $0.01$) was sufficient. When the system is fully observed, the observed states $\mathbf{u}$ can be used directly.

\section{Experiments}
\label{sec:experiments}

We benchmarked our curriculum multiple shooting (CMS) approach in two settings: (i) fitting UDEs and NODEs, and (ii) fitting mechanistic ODE models to noisy time-series data. The UDE and NODE benchmarks comprised three models of varying complexity and covered both simulated (one model) and real (two models) time-series data. The mechanistic ODE benchmarks provided a complementary robustness test and comprised eight real-data models from the PEtab benchmark collection (Tab.~\ref{tab:mechanistic_models})~\citep{hass_benchmark_2019, schmiester_petabinteroperable_2021}.

All UDE/NODE benchmarks were implemented in Python using the JAX-based packages \texttt{equinox} and \texttt{diffrax}~\citep{kidger_neural_2022}, whereas the mechanistic benchmarks were implemented in Julia using \texttt{PEtab.jl}~\citep{persson_petabjl_2025}. All experiments were run on CPUs, with a single model fit typically requiring less than 15 minutes (App.~\ref{app:experiment_details}). Code for all benchmarks is provided in the attached zip file and will be made available in a public repository upon de-anonymization.

\subsection{Models and datasets}
\label{sec:models_datasets}

We briefly summarize the UDE and mechanistic ODE models below; full details on model architectures and datasets are provided in Appendix~\ref{app:benchmark_details}.

\textbf{Lotka--Volterra UDE and NODE.}
The Lotka--Volterra system models prey ($x$) and predator ($y$) populations in ecology and exhibits oscillatory dynamics, making it challenging to fit~\citep{pitt_parameter_2019}. We used simulated noise-corrupted data as in~\citep{rackauckas_universal_2021, ko_homotopy-based_2023}, with $x$ and $y$ observed over one period. We considered two model architectures. The first, following~\citep{rackauckas_universal_2021}, is a UDE of the form $\mathrm{d}x / \mathrm{d}t = \alpha x + \mathbf{NN}_1(x,y;\bm{\theta}_n)$ and $\mathrm{d}y / \mathrm{d}t = -\delta y + \mathbf{NN}_2(x,y;\bm{\theta}_n)$, where $\mathbf{NN}$ is a feed-forward neural network and $\bm{\theta}_m = (\alpha,\delta)$ are mechanistic parameters. We jointly learned the neural-network parameters $\bm{\theta}_n$ and the mechanistic parameters $(\alpha,\delta)$. To discourage the network from absorbing the full dynamics and driving $(\alpha,\delta)$ toward zero, we applied output regularization (Eq.~\ref{eq:output_reg})~\citep{philipps_non-negative_2024}. Second, we considered a pure NODE structure (Eq.~\ref{eq:lv_node_eq}).

\textbf{SIR UDE.}
We used an SIR (susceptible, infected, and recovered) UDE model of the initial COVID-19 outbreak in the United Kingdom (UK), as described in~\citep{dandekar_machine_2020}. The model augments a classical SIR system with a feed-forward neural network to capture quarantine strength (Eq.~\ref{eq:sir_eq}). We fitted the model to real time-series data from the UK. The model is partially observed, since only the infected and recovered populations are measured, and it exhibits pronounced scale differences across states.

\textbf{Double pendulum UDE.}
We used real data from~\citep{schmidt_distilling_2009} describing the dynamics of a hard-to-fit double-pendulum system. As in~\citep{ko_homotopy-based_2023}, we employed a physics-informed second-order UDE (Eq~\ref{eq:pendulum_ude}).

\textbf{Mechanistic ODE models.}
We used eight mechanistic models from the PEtab benchmark collection~\citep{hass_benchmark_2019, schmiester_petabinteroperable_2021}. These models span diverse biological applications (e.g. cell signaling, gene regulation, and disease spread), and are all paired with real data. They cover a wide range of challenges for model fitting, including stiffness, partial observability, model misspecification, irregular sampling, variation in model size, and changing simulation conditions across measurements (Tab.~\ref{tab:mechanistic_models}).

\subsection{Evaluation metrics and baselines}
\label{sec:eval_metrics}

We briefly summarize the evaluation metrics and baselines used in the experiments. Additional implementation details and hyperparameter tuning procedure are provided in App.~\ref{app:experiment_details}.

\textbf{UDE and NODE models.}
For evaluation, we split the time series into training and validation sets (50/50 for Lotka--Volterra and 70/30 for the other models), and tuned hyperparameters on the training data. Performance was evaluated using mean-squared error (MSE) on each set, hereafter referred to as interpolation and extrapolation MSE, following~\citep{ko_homotopy-based_2023}. MSE was reported at the end of training, with all strategies run for the same number of epochs per model. To account for sensitivity to initialization, each experiment was repeated over 100 random initializations of the neural-network and mechanistic parameters. As some training runs diverged and produced skewed MSE distributions with outliers (Fig.~\ref{fig:ude_benchmarks}), we summarized performance using the median MSE and compared methods using the Wilcoxon rank-sum test, with per-model Holm-corrected p-values~\citep{holm_simple_1979}. Tables of values are provided in App~\ref{app:additional_results}. In figures, we used ns ($p \geq 0.05$), * ($p < 0.05$), ** ($p < 0.01$), and *** ($p < 0.001$).

For \textbf{baselines}, because our experiments include both fully and partially observed models, we compared against four state-of-the-art methods applicable to both settings. These where single-shooting training with the Adam optimizer (Adam)~\citep{kingma_adam_2017}, curriculum learning with Adam (CL; Sec.~\ref{sec:curriculum_learning}), multiple shooting (MS; Sec.~\ref{sec:multiple_shooting}), and Adam followed by a second-order optimizer with a BFGS Hessian approximation (Adam+BFGS). The latter is a common strategy for training both UDEs and PINNs~\citep{rackauckas_universal_2021, rathore_challenges_2024}. Results were reported for the best-performing hyperparameter configuration for each training strategy (Tabs.~\ref{tab:adam_tuning}--\ref{tab:cl_ms_tuning}).

\textbf{Mechanistic ODE models.}
For evaluation, we counted how many of 1000 training runs from randomly sampled parameter initializations converge to the global minimum (Fig.~\ref{fig:mech_benchmarks}a), a standard metric in systems biology~\citep{hass_benchmark_2019, raue_lessons_2013}. Since second-order optimizers are effective for mechanistic ODEs~\citep{raue_lessons_2013}, we use the Fides trust-region Newton optimizer~\citep{frohlich_fides_2022}. We compare CMS against three baselines: plain Fides (control), curriculum learning with Fides (CL), and multiple shooting with Fides (MS). Because our goal was to assess robustness of our CMS strategy rather than maximize performance, we used a small fixed grid of relevant hyperparameter values and report the best-performing combinations (Tab.~\ref{tab:mech_tuning}).

\subsection{Results on UDE and NODE benchmarks}
\label{sec:ude_node_benchmarks}

\begin{figure}[h!]
    \centering
    \includegraphics[width=0.95\linewidth]{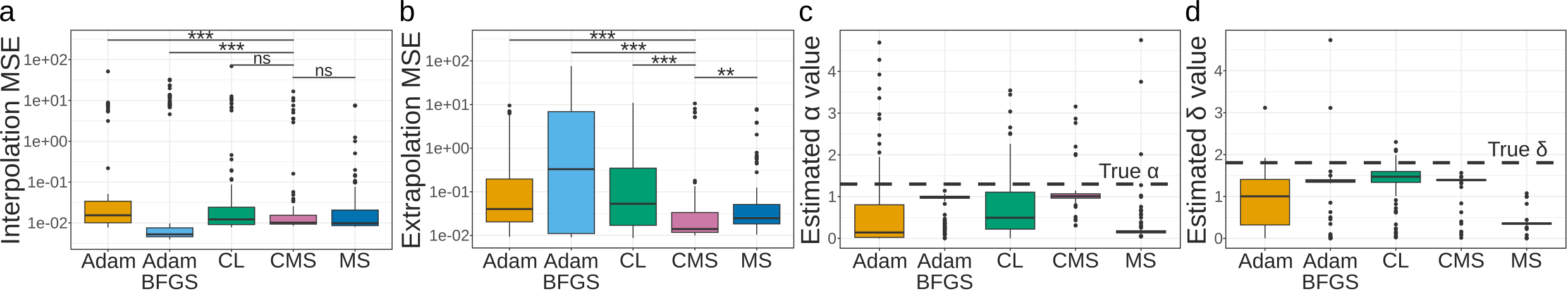}
    \caption{Benchmark results for the Lotka--Volterra UDE model from 100 multistarts with random parameter initializations. (a,b) Interpolation and extrapolation MSE for each strategy on $\log_{10}$ scale; p-values reported using the Wilcoxon test. (c,d) Estimated values of the mechanistic parameters.}
    \label{fig:lv_ude}
\end{figure}

For the Lotka--Volterra UDE with simulated data, CMS achieved the lowest median extrapolation MSE, with a value of 0.0148 compared to 0.0251 for the second-best method, multiple shooting (Fig.~\ref{fig:lv_ude}b; $p = \num{1.3e-3}$). For interpolation MSE (Fig.~\ref{fig:lv_ude}a), CMS was on par with both curriculum learning (0.010 vs.\ 0.012; $p = 0.37$) and multiple shooting (0.010 vs.\ 0.0099; $p = 0.12$). Adam+BFGS achieved the lowest interpolation MSE (0.0053; $p = \num{1e-9}$), but performed substantially worse than CMS in extrapolation (0.0148 vs.\ 1.52; $p = \num{7e-4}$). Inspection of the fitted training trajectories showed that Adam+BFGS overfitted the training data noise (Fig.~\ref{fig:bfgs_overfit}), explaining its worse extrapolation error. As a robustness check, we verified that CMS remained top-performing under higher measurement noise and longer simulation horizons (Fig.~\ref{fig:lv_ude_noise}).

Strikingly, CMS also recovered the Lotka--Volterra UDE mechanistic parameters close to the ground truth (true $(\alpha,\delta)=(1.3,1.8)$, fitted median $(1.0,1.45)$). In contrast, multiple shooting drove both parameters toward zero (Fig.~\ref{fig:lv_ude}c-d), indicating that the UDE degenerated to a NODE. This was likely because CMS could be trained with stronger output regularization (Eq.~\ref{eq:output_reg}) without increasing the interpolation MSE (Fig.~\ref{fig:output_reg}). While too strong regularization eventually degraded performance for all strategies (Fig.~\ref{fig:output_reg}), CMS was more robust, indicating that it is better suited to harder optimization objectives like those induced by strong regularization.

\begin{figure}[ht]
    \centering
    \includegraphics[width=0.99\linewidth]{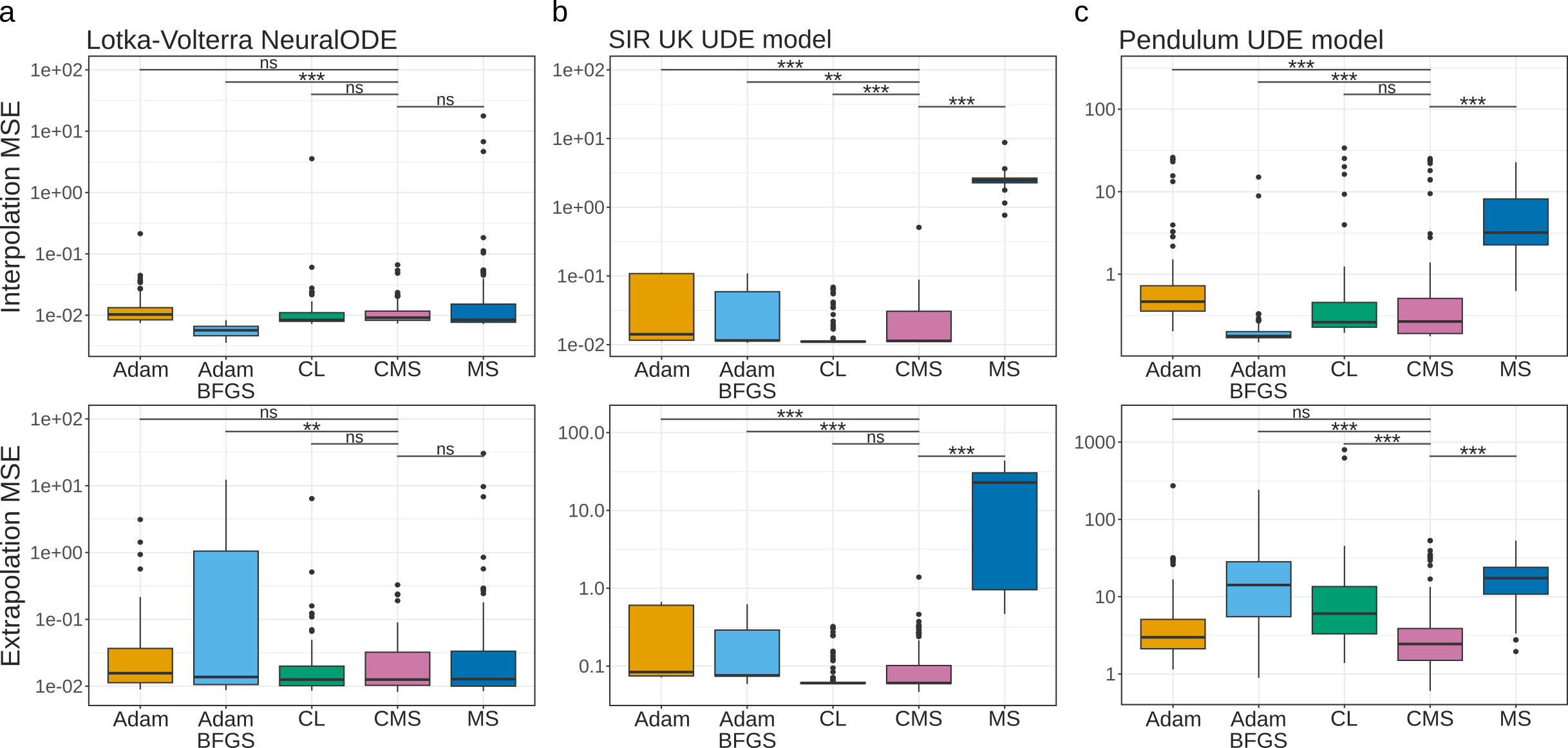}
    \caption{Benchmark UDE results from 100 multistarts with random parameter initializations. Interpolation MSE (top) and extrapolation MSE (bottom) on $\log_{10}$ scale for the (a) Lotka--Volterra NODE, (b) SIR UDE, and (c) double pendulum models. P-values reported using the Wilcoxon test.}
    \label{fig:ude_benchmarks}
\end{figure}

Across the remaining NODE and UDE models, CMS was top-performing in extrapolation MSE (Fig.~\ref{fig:ude_benchmarks}). For the Lotka--Volterra NODE (Fig.~\ref{fig:ude_benchmarks}a), CMS matched the best method, curriculum learning (CL) (0.0126 vs.\ 0.0126; $p = 0.99$). Adam+BFGS again extrapolated worse than CMS (0.0126 vs.\ 0.0180; $p = \num{6.8e-3}$), consistent with its lower interpolation MSE reflecting overfitting. For the SIR UDE (Fig.~\ref{fig:ude_benchmarks}b), CMS again matched the best method (CL) (0.0606 vs.\ 0.0605; $p = 0.47$), while outperforming the second-best (Adam+BFGS) (0.0606 vs.\ 0.0759; $p = \num{7e-11}$). For the double pendulum (Fig.~\ref{fig:ude_benchmarks}c), CMS extrapolated best, slightly outperforming the second best, Adam, in extrapolation MSE (2.45 vs.\ 2.99; $p = 0.058$), and clearly in interpolation MSE (0.267 vs.\ 0.466; $p = \num{6e-6}$). As CMS was designed to improve upon multiple shooting, we note it substantially improved extrapolation MSE over multiple shooting for the SIR (0.0606 vs.\ 24.9; $p = \num{5e-33}$) and the pendulum models (2.45 vs.\ 17.6; $p = \num{7e-19}$). On the Lotka--Volterra NODE and UDE models, where multiple shooting performed well (Figs.~\ref{fig:lv_ude}--\ref{fig:ude_benchmarks}), CMS was also more robust to hyperparameter choice, particularly the continuity penalty $\lambda_{\mathrm{cont}}$ (Fig.~\ref{fig:sense_ude}). This robustness also transferred across implementations and gradient-computation schemes, with CMS performing well on the Lotka--Volterra models and SIR UDE in complementary Julia-based benchmarks (App.~\ref{sec:julia_ude_benchmark}).

To assess epoch-wise convergence, we further inspected single-shooting interpolation-MSE curves over epochs (Fig.~\ref{fig:training_traces}). CMS reached low MSE in fewer epochs than the other strategies on $3/4$ models. The only exception is the non-oscillatory SIR model, where CL converged faster. This is discussed in more detail in App.~\ref{app:training_traces}.

In summary, CMS was consistently top-performing in terms of both interpolation and extrapolation MSE, while typically being the fastest to converge during training.

\subsection{Results on mechanistic ODE benchmarks}

\begin{figure}[ht]
    \centering
    \includegraphics[width=0.99\linewidth]{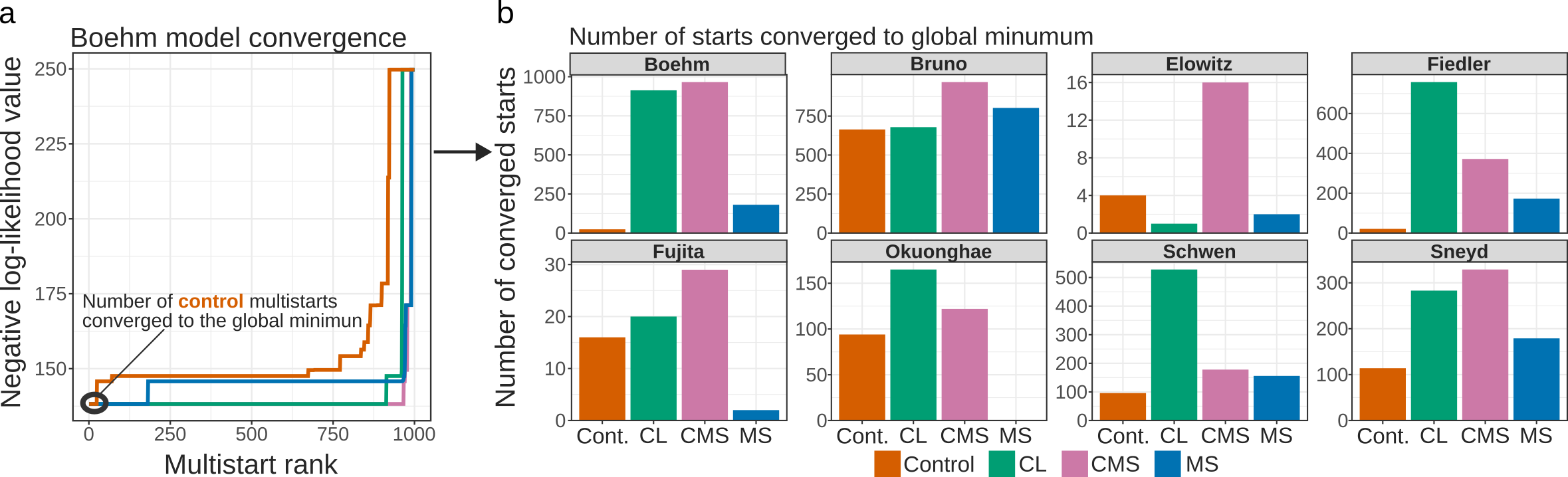}
    \caption{Benchmark results on eight mechanistic ODE models from 1000 multistarts with random parameter initializations. (a) Illustration of the evaluation criterion, i.e. the number of runs converging to the global optimum. (b) Results for each model. Control denotes plain single-shooting optimization.}
    \label{fig:mech_benchmarks}
\end{figure}

To assess the robustness of CMS more broadly, we evaluated it on eight diverse mechanistic ODE models with features that make model fitting challenging (e.g. stiffness, partial observability; Tab.~\ref{tab:mechanistic_models}). Using the number of runs, out of 1000 random initialisations, converging to the global optimum as the metric (Fig.~\ref{fig:mech_benchmarks}a), CMS outperformed the single-shooting control on all models by an average fold-change of 8.9 (range: 1.3--40.2). CMS also outperformed multiple shooting on all models and was the best-performing method on 5/8 models (Fig.~\ref{fig:mech_benchmarks}b). CL outperformed the single-shooting control on 7/8 models and was best on the remaining 3/8 models. As all methods used the same small hyperparameter grid, the relative performance of CL and CMS may partly reflect tuning difficulty, since CL has fewer hyperparameters to tune.

Because CL performed well on a subset of benchmarks and has fewer tuning parameters than CMS, making it attractive when effective, we next investigated under which conditions it worked well. Although no single definitive pattern emerged, the results suggested that CL performed best when early curriculum stages were representative of the full problem, for example when each curriculum stage contained all observables (i.e. all model outputs). This is discussed in more detail in Appendix~\ref{app:mech_benchmarks}.

\subsection{Why does curriculum multiple shooting improve training?}
\label{sec:hess_loss_landscapes}

\begin{figure}[ht]
    \centering
    \includegraphics[width=0.99\linewidth]{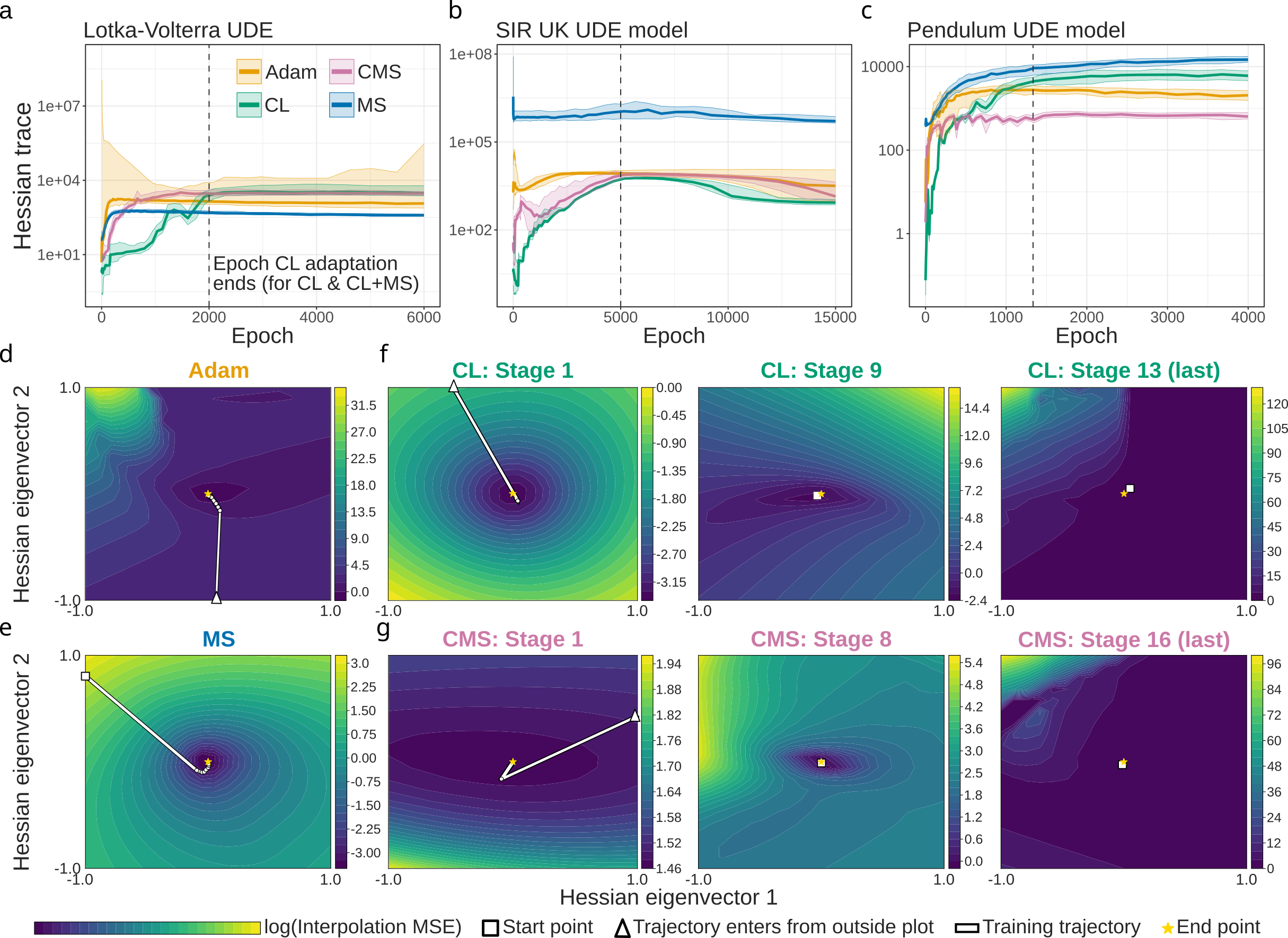}
    \caption{Training trajectory analysis across training strategies. (a--c) Hessian trace profiles for the Lotka--Volterra UDE, SIR UDE, and double pendulum UDE models. Data are shown as median (solid lines) and 25th--75th percentiles (shaded intervals) across 100 multistarts. (d--g) Loss-landscape visualisations along the top two Hessian eigenvectors for the Lotka--Volterra UDE, one representative multistart per strategy reaching interpolation MSE $< 0.02$. For Adam and MS, the Hessian eigenvectors are computed at the final epoch; for CL and CMS at the final epoch of each curriculum stage.}
    \label{fig:hess_trace}
\end{figure}

To gain intuition for the strong performance of CMS, we analysed the training trajectories on the UDE/NODE models using Hessian trace profiles~\citep{yao_pyhessian_2020}. Small, positive Hessian traces indicate good conditioning of the optimisation problem and are considered favourable for training \citep{ko_homotopy-based_2023, yao_pyhessian_2020}. CL and CMS consistently showed lower Hessian traces than Adam and MS early in training (Fig.~\ref{fig:hess_trace}a--c), often followed by a gradual increase, consistent with the expected progressive learning behaviour of curriculum approaches. This was also reflected in gradient $\ell_2$-norm traces (Fig.~\ref{fig:grad_magnitude}): CL and CMS kept gradient magnitudes controlled throughout training, with gradients decreasing within stages and increasing at curriculum stage boundaries. By contrast, poor training performance for Adam and MS coincided with larger gradient norms. Large gradients for Adam may result from long time spans for numerical integration by the ODE solver, which, building on the interpretation of NODEs as continuous-depth residual neural networks~\citep{chen_neural_2018,massaroli_dissecting_2020}, can make gradients more prone to vanishing or exploding~\citep{pascanu_difficulty_2013}. While CL, MS, and CMS all initially train on shorter trajectories, MS additionally relies on a strong continuity penalty to make its windowed loss an effective surrogate for single shooting training, which may introduce large gradients.

Motivated by evidence that skip-connections can convexify loss landscapes of residual neural networks by reducing the effective depth~\citep{li_visualizing_2018}, we also analysed the loss landscape of the Lotka--Volterra UDE along the top two Hessian eigenvectors, evaluated at the final epoch of stages or overall training (Fig.~\ref{fig:hess_trace}d--g and Fig.~\ref{fig:loss_landscape_si}). Consistent with~\citep{ko_homotopy-based_2023}, Adam exhibited a rugged, non-convex landscape (Fig.~\ref{fig:hess_trace}d), whereas MS (Fig.~\ref{fig:hess_trace}e) and early stages of CL and CMS (Fig.~\ref{fig:hess_trace}f--g) showed smoother, more convex landscapes. Unlike MS, which never transitions to the full single-shooting objective (Eq.~\ref{eq:objective}), CL and CMS eventually progressed to more complex landscapes with multiple plateaus and local minima (see also Fig.~\ref{fig:loss_landscape_si} ). However, earlier stages often positioned optimisation in a favourable region before the harder landscape was encountered, as reflected by the small displacement between starting (white square) and final (yellow star) points in later stages. This was particularly evident for CMS, which is exposed to the full trajectory from the start, whereas CL introduces additional training data abruptly, which can lead to larger shifts between stages (Fig.~\ref{fig:loss_landscape_si}c). Analysis of single-shooting interpolation MSE traces revealed non-monotonic progression in both CL and CMS (Fig.~\ref{fig:training_traces}). While such non-monotonicity may reflect a desirable ability to escape local minima, the larger spikes in CL suggest that its early stages are often not representative of the single-shooting objective.

In summary, CMS likely achieves strong performance by easing early-phase optimisation while remaining aligned with the single-shooting objective, ultimately recovering it and reducing reliance on a strong continuity penalty.

\section{Discussion and limitations}
\label{sec:discussion}

We propose \emph{curriculum multiple shooting} (CMS), a general-purpose training strategy for fitting mechanistic ODEs, NODEs, and hybrid UDEs to noisy time-series data. Across diverse benchmark problems spanning full and partial observability, noisy and irregularly sampled measurements, and challenging dynamics, CMS was consistently among the top-performing methods (Fig.~\ref{fig:lv_ude}--\ref{fig:mech_benchmarks}). Its strong performance on classical mechanistic ODE fitting tasks (Fig.~\ref{fig:mech_benchmarks}) further demonstrates that our results extend beyond ML to disciplines where mechanistic modelling remains central, such as biology, epidemiology, and pharmacology. More broadly, this suggests that training NODEs and UDEs is fundamentally a hard ODE fitting problem, and that training strategies can transfer from mechanistic ODEs to NODEs/UDEs, and likely also to related discrete-time nonlinear state-space models where multiple-shooting ideas have been used~\citep{ribeiro_smoothness_2020}.

Our study also has several limitations pointing to relevant future directions. First, we consider only a limited number of NODE and UDE benchmarks, partly because identifying and implementing suitable benchmark problems is labour-intensive. This underscores the need for broader, standardized benchmark collections similar to those available for mechanistic ODEs~\citep{hass_benchmark_2019}. Second, the models considered here are low-dimensional ($<20$ states), so it remains unclear how CMS scales to high-dimensional data, such as omics time-series data that are increasingly modeled with NODEs in biology~\citep{hossain_biologically_2024, richter_generative_2026}. Third, CMS introduces several tuning parameters. Improving default heuristics and developing robust tuning schemes would further increase its practical usability.

\newpage

\begin{ack}
FF, GF, BS and SP were supported by the Francis Crick Institute, which receives its core
funding from Cancer Research UK (CC2242), the UK Medical Research Council
(CC2242), and the Wellcome Trust (CC2242), as well as the European Union (ERC,
DeepMechanism, grant no 101163005).
\end{ack}

\bibliographystyle{unsrtnat}
\bibliography{UDE_training}

\newpage

\appendix
\renewcommand{\thefigure}{\thesection.\arabic{figure}}
\renewcommand{\thetable}{\thesection.\arabic{table}}
\renewcommand{\theequation}{\thesection.\arabic{equation}}

\renewcommand{\theHfigure}{appendix.\thesection.\arabic{figure}}
\renewcommand{\theHtable}{appendix.\thesection.\arabic{table}}
\renewcommand{\theHequation}{appendix.\thesection.\arabic{equation}}

\setcounter{figure}{0}
\setcounter{table}{0}
\setcounter{equation}{0}

\section{Additional Results}
\label{app:additional_results}

This section introduces additional results, figures, and tables not included in the main text for brevity. It also contains results from a robustness benchmark on UDEs in Julia, complementing the Python based benchmarks in Sec.~\ref{sec:ude_node_benchmarks}.

\subsection{Summary table of benchmark results}

This section contains a table summarizing the benchmark results for the NODE and UDE models in Sec.~\ref{sec:ude_node_benchmarks} in Tabs.~\ref{tab:lv_ude_res}--\ref{tab:pendulum_res}.

\begin{table}[ht]
    \centering
    \caption{Benchmark results for the Lotka--Volterra UDE (Eq.~\ref{eq:lv_ude_eq}). Interpolation and extrapolation MSE across all 100 multi-starts are reported as $\mathrm{median} \pm \mathrm{median\ absolute\ deviation}$ for each training strategy. The best interpolation and extrapolation values are shown in bold.}
    \label{tab:lv_ude_res}
    \begin{tabular}{lcc}
        \toprule
        Strategy & Interpolation MSE & Extrapolation MSE \\
        \midrule
        CMS & $0.0101 \pm 0.00173$ & $\mathbf{0.0148 \pm 0.00610}$ \\
        CL & $0.0123 \pm 0.00551$ & $0.0785 \pm 0.0996$ \\
        MS & $0.00990 \pm 0.00235$ & $0.0251\pm 0.0134$ \\
        Adam & $0.0242 \pm 0.0226$ & $0.188 \pm 0.262$ \\
        Adam+BFGS & $\mathbf{0.00526 \pm 0.00131}$ & $1.52 \pm 2.23$ \\
        \bottomrule
    \end{tabular}
\end{table}

\begin{table}[ht]
    \centering
    \caption{Benchmark results for the Lotka--Volterra NODE (Eq.~\ref{eq:lv_node_eq}). Values are reported as in Tab.~\ref{tab:lv_ude_res}.}
    \label{tab:lv_node_res}
    \begin{tabular}{lcc}
        \toprule
        Strategy & Interpolation MSE & Extrapolation MSE \\
        \midrule
        CMS & $0.00913 \pm 0.00165$ & $0.0126 \pm 0.00476$ \\
        CL & $0.00837 \pm 0.000975$ & $\mathbf{0.0126 \pm 0.00410}$ \\
        MS & $0.00841 \pm 0.00134$ & $0.0128 \pm 0.00484$ \\
        Adam & $0.0103 \pm 0.00318$ & $0.0157 \pm 0.00794$ \\
        Adam+BFGS & $\mathbf{0.00569 \pm 0.00145}$ & $0.0180 \pm 0.0135$ \\
        \bottomrule
    \end{tabular}
\end{table}

\begin{table}[ht]
    \centering
    \caption{Benchmark results for the SIR (Eq.~\ref{eq:sir_eq}). Values are reported as in Tab.~\ref{tab:lv_ude_res}.}
    \label{tab:sir_res}
    \begin{tabular}{lcc}
        \toprule
        Strategy & Interpolation MSE & Extrapolation MSE \\
        \midrule
        CMS & $0.0113 \pm 0.000502$ & $0.0606 \pm 0.00378$ \\
        CL & $\mathbf{0.0111 \pm 0.000166}$ & $\mathbf{0.0605 \pm 0.00137}$ \\
        MS & $13.3 \pm 15.3$ & $24.9 \pm 29.1$ \\
        Adam & $0.0142 \pm 0.00437$ & $0.0839 \pm 0.0175$ \\
        Adam+BFGS & $0.0115 \pm 0.000321$ & $0.0759 \pm 0.00373$ \\
        \bottomrule
    \end{tabular}
\end{table}

\begin{table}[h!]
    \centering
    \caption{Benchmark results for the double pendulum model (Eq.~\ref{eq:pendulum_ude}). Values are reported as in Tab.~\ref{tab:lv_ude_res}.}
    \label{tab:pendulum_res}
    \begin{tabular}{lcc}
        \toprule
        Strategy & Interpolation MSE & Extrapolation MSE \\
        \midrule
        CMS & $0.267 \pm 0.127$ & $\mathbf{2.45 \pm 1.66}$ \\
        CL & $0.265 \pm 0.0687$ & $6.07 \pm 5.16$ \\
        MS & $3.19 \pm 2.31$ & $17.6 \pm 10.1$ \\
        Adam & $0.466 \pm 0.204$ & $2.99 \pm 1.69$ \\
        Adam+BFGS & $\mathbf{0.178 \pm 0.0167}$ & $15.7 \pm 16.5$ \\
        \bottomrule
    \end{tabular}
\end{table}

\newpage

\subsection{Additional Figures}

\begin{figure}[ht]
    \centering
    \includegraphics[width=0.85\linewidth]{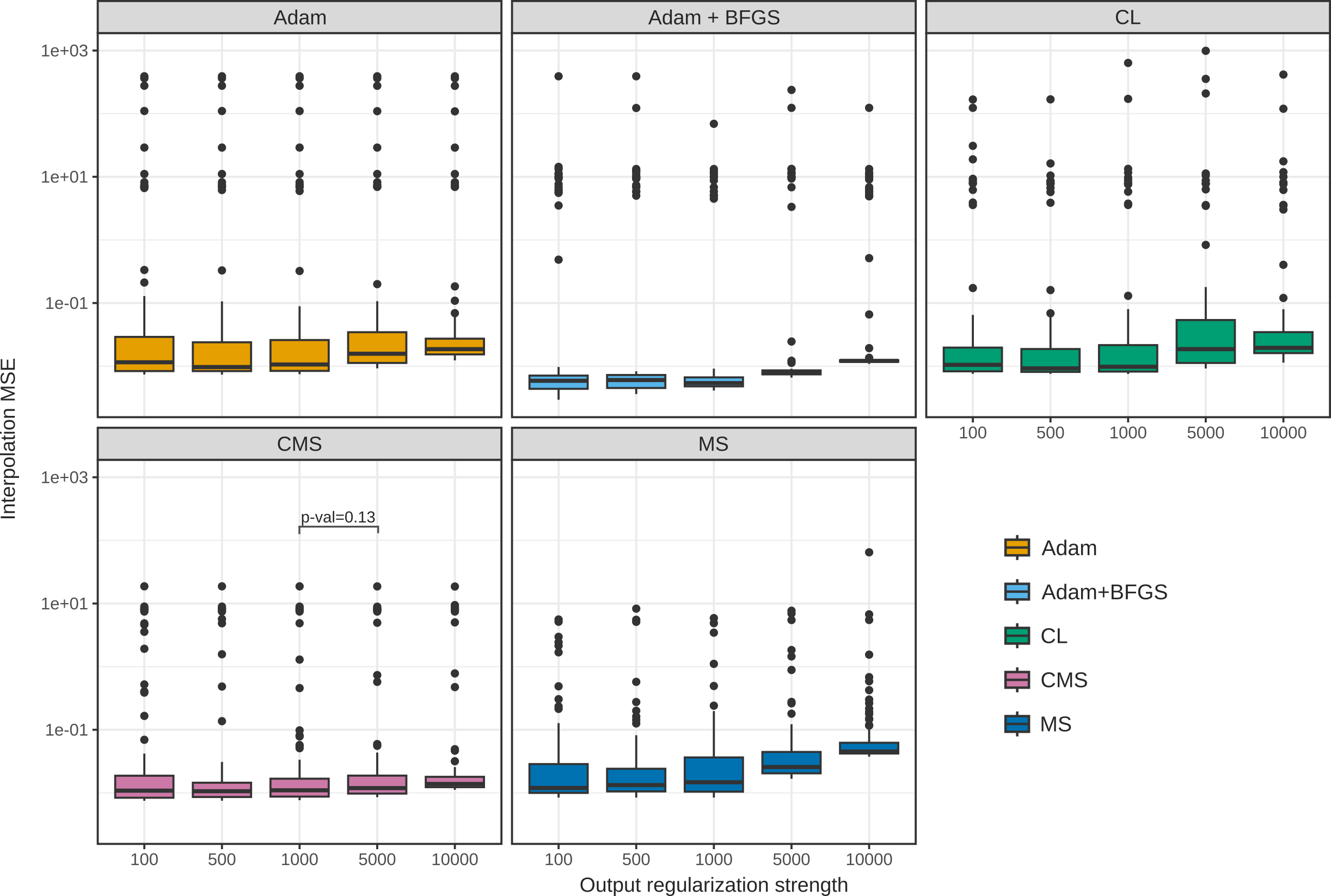}
    \caption{Effect of increasing output regularization strength ($\lambda_O$ in Eq.~\ref{eq:output_reg}) on the interpolation MSE of the Lotka--Volterra UDE model for each training strategy. Results are based on 100 multistarts with random parameter initializations and are shown on a $\log_{10}$ scale. P-values are reported using the Wilcoxon rank-sum test. The tested $\lambda_O$ values are a subset of those considered during hyperparameter tuning (Tabs.~\ref{tab:ms_tuning}--\ref{tab:cl_ms_tuning}).}
    \label{fig:output_reg}
\end{figure}

Fig.~\ref{fig:output_reg} shows the effect of increasing output regularization strength ($\lambda_O$, Eq.~\ref{eq:output_reg}) on the training/interpolation MSE of the Lotka--Volterra UDE model. Although too strong regularization ultimately degrades training performance for all strategies, curriculum multiple shooting (CMS) is the most robust. This suggests that CMS is better suited to harder optimization objectives, such as those induced by stronger regularization.

\begin{figure}[h]
    \centering
    \includegraphics[width=0.75\linewidth]{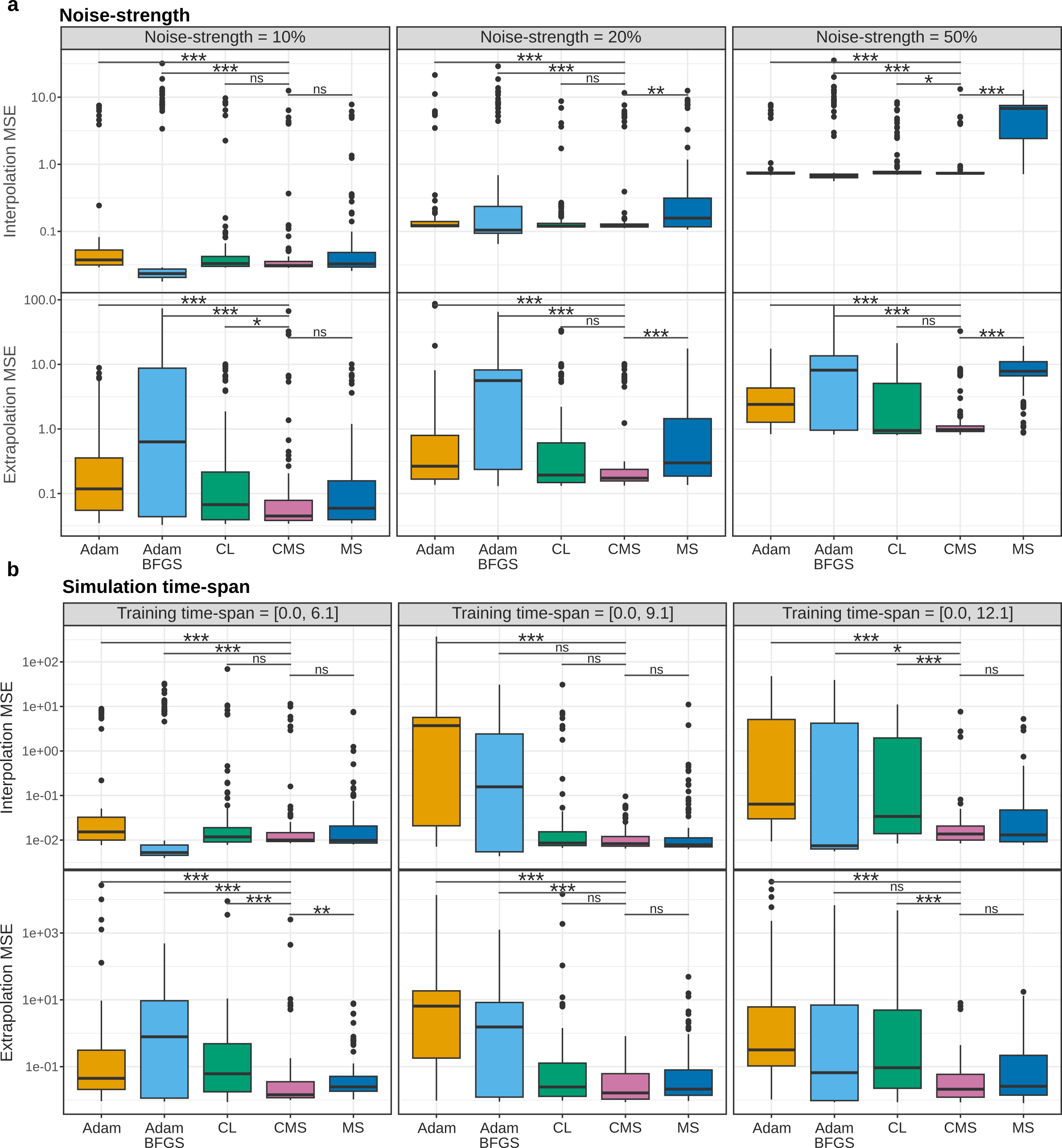}
    \caption{Effect of increasing (a) measurement noise and (b) simulation time-span on the Lotka--Volterra UDE model for each training strategy, evaluated using interpolation MSE (top row) and extrapolation MSE (bottom row). Results are based on 100 multistarts with random parameter initializations and are shown on a $\log_{10}$ scale. Holm-corrected p-values per model are reported using the Wilcoxon rank-sum test, with ns ($p \geq 0.05$), * ($p < 0.05$), ** ($p < 0.01$), and *** ($p < 0.001$). Due to computational cost, the same hyperparameter values as in the benchmark in Fig.~\ref{fig:lv_ude} were used.}
    \label{fig:lv_ude_noise}
\end{figure}

Fig.~\ref{fig:lv_ude_noise} provides robustness checks for CMS on the Lotka--Volterra UDE under higher measurement noise and longer simulation horizons. The strong performance of CMS, particularly for longer time spans, further supports its suitability for training on longer time-series data, which have been shown to make NODE and UDE training more difficult~\citep{ko_homotopy-based_2023}.

\begin{figure}[h]
    \centering
    \includegraphics[width=0.75\linewidth]{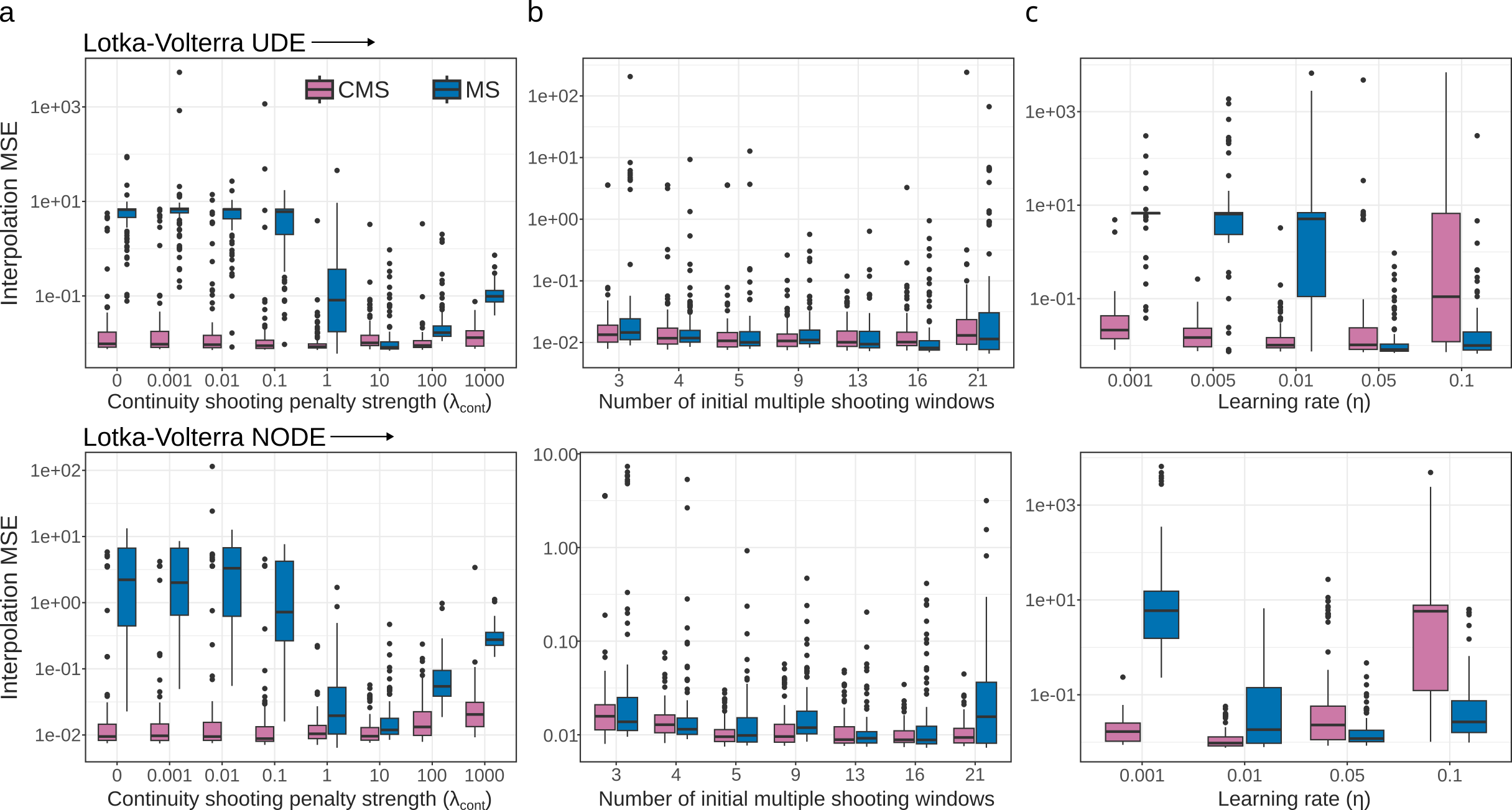}
    \caption{Sensitivity of interpolation MSE to the hyperparameters (a) continuity penalty ($\lambda_{\mathrm{cont}}$ in Eq.~\ref{eq:ms_penalty}), (b) number of initial multiple shooting windows shooting windows, and (c) learning rate for the multiple shooting (MS) and curriculum multiple shooting (CMS) training strategies on the Lotka--Volterra UDE (top row) and Lotka--Volterra NODE (bottom row) models. Results are based on 100 multistarts with random parameter initializations and are shown on a $\log_{10}$ scale. The tested values are those considered during hyperparameter tuning Tabs.~\ref{tab:ms_tuning}--~\ref{tab:cl_ms_tuning}.}
    \label{fig:sense_ude}
\end{figure}

Fig.~\ref{fig:sense_ude} investigates the sensitivity of multiple shooting (MS) and CMS to hyperparameters on the two models where MS performed well. In particular, for the hard-to-tune continuity penalty $\lambda_{\mathrm{cont}}$ in Eq.~\ref{eq:ms_penalty}, CMS is substantially more robust, achieving lower interpolation MSE over a wider range of values.

\begin{figure}[ht]
    \centering
    \includegraphics[width=0.80\linewidth]{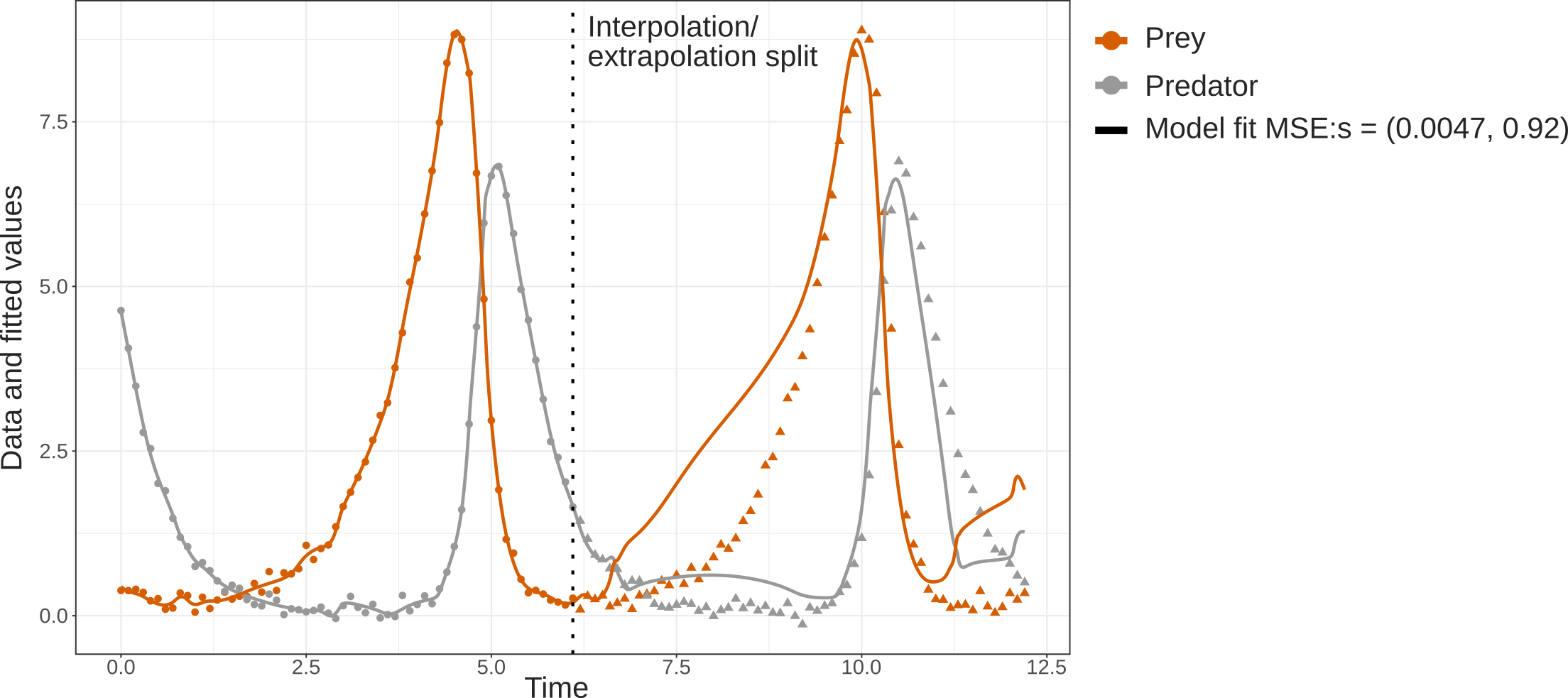}
    \caption{Adam+BFGS overfits the Lotka--Volterra UDE training data for the prey and predator observables. Lines shows a representative model fit with low interpolation/training error but high extrapolation/validation error. MSE values are reported in the legend as (interpolation, extrapolation).}
    \label{fig:bfgs_overfit}
\end{figure}

Fig.~\ref{fig:bfgs_overfit} shows an example where the Adam+BFGS training strategy overfits the Lotka--Volterra UDE model, achieving low interpolation MSE but high extrapolation MSE.

\begin{figure}[ht]
    \centering
    \includegraphics[width=0.99\linewidth]{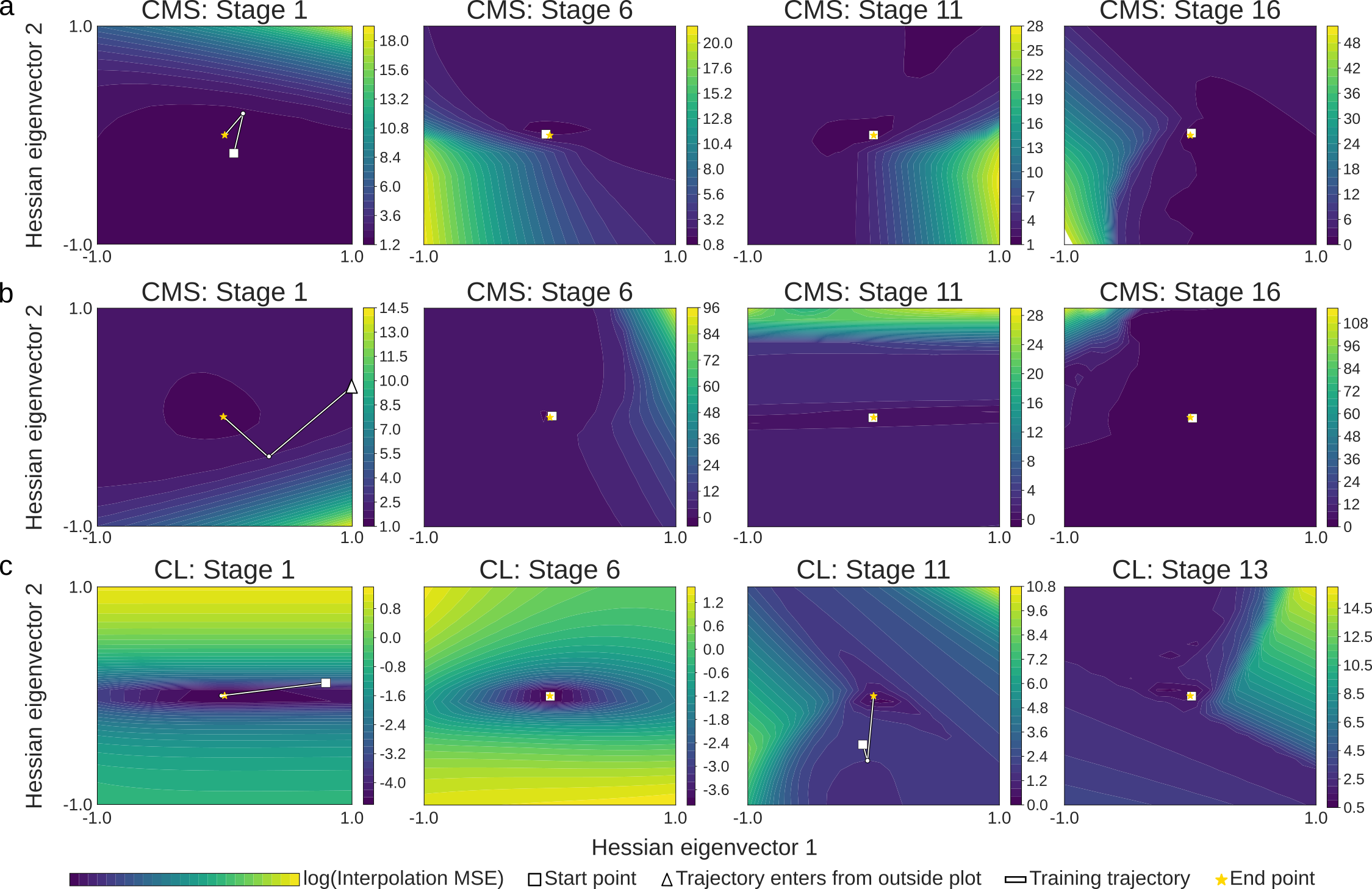}
    \caption{Loss-landscape visualisations for representative multistarts of the Lotka--Volterra UDE with interpolation MSE $<0.02$, shown along the top two Hessian eigenvectors. Each panel shows one representative multistart for (a--b) curriculum multiple shooting and (c) curriculum learning. Landscapes are evaluated at the final epoch of the corresponding curriculum stages.}
    \label{fig:loss_landscape_si}
\end{figure}

Finally, Fig.~\ref{fig:loss_landscape_si} provides additional loss-landscape visualizations. Overall, these plots show that in later curriculum stages, where the objective becomes harder to optimize, CMS is typically already positioned in a favourable region of the loss landscape. This is reflected by the small displacement between the starting point (white square/triangles) and endpoint (yellow star) within later stages.

\clearpage

\begin{figure}[ht]
    \centering
    \includegraphics[width=0.99\linewidth]{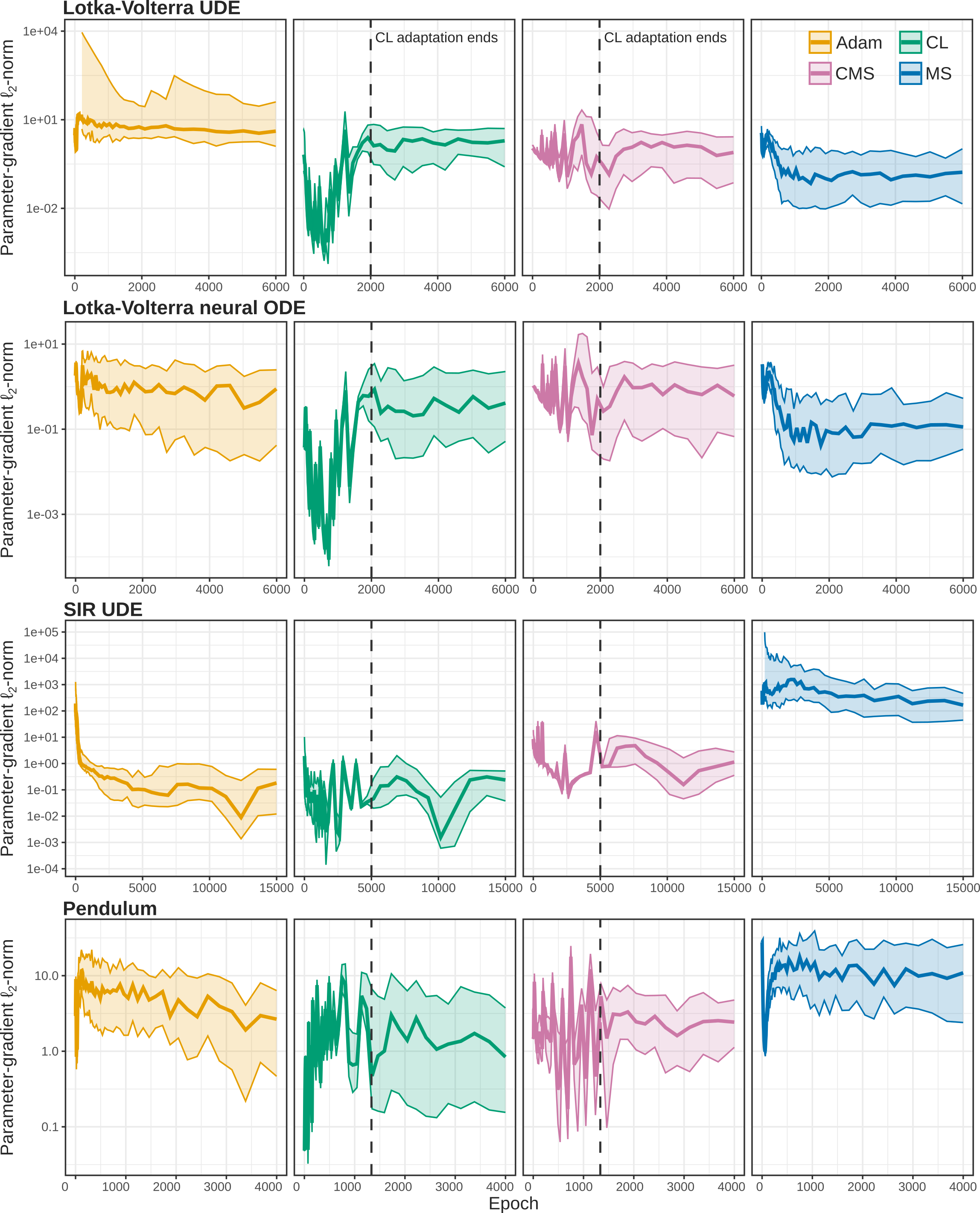}
    \caption{$\ell_2$ norm of the parameter (both neural and mechanistic) gradient of the training objective (e.g. for MS this corresponds to the multiple-shooting objective in Eq.~\ref{eq:ms_objective}) for the Lotka--Volterra UDE (top), Lotka--Volterra NODE (upper middle), SIR UDE (lower middle), and pendulum UDE (bottom) models. Results are based on $100$ multistarts with random parameter initializations and are shown on a $\log_{10}$ scale. Solid lines denote the median, and shaded intervals denote the 25th--75th percentiles across multistarts. Dashed vertical lines indicate the end of curriculum adaptation for the relevant strategies.}
    \label{fig:grad_magnitude}
\end{figure}

\subsection{Additional analysis on UDE/NODE training traces}
\label{app:training_traces}

\begin{figure}[h]
    \centering
    \includegraphics[width=0.99\linewidth]{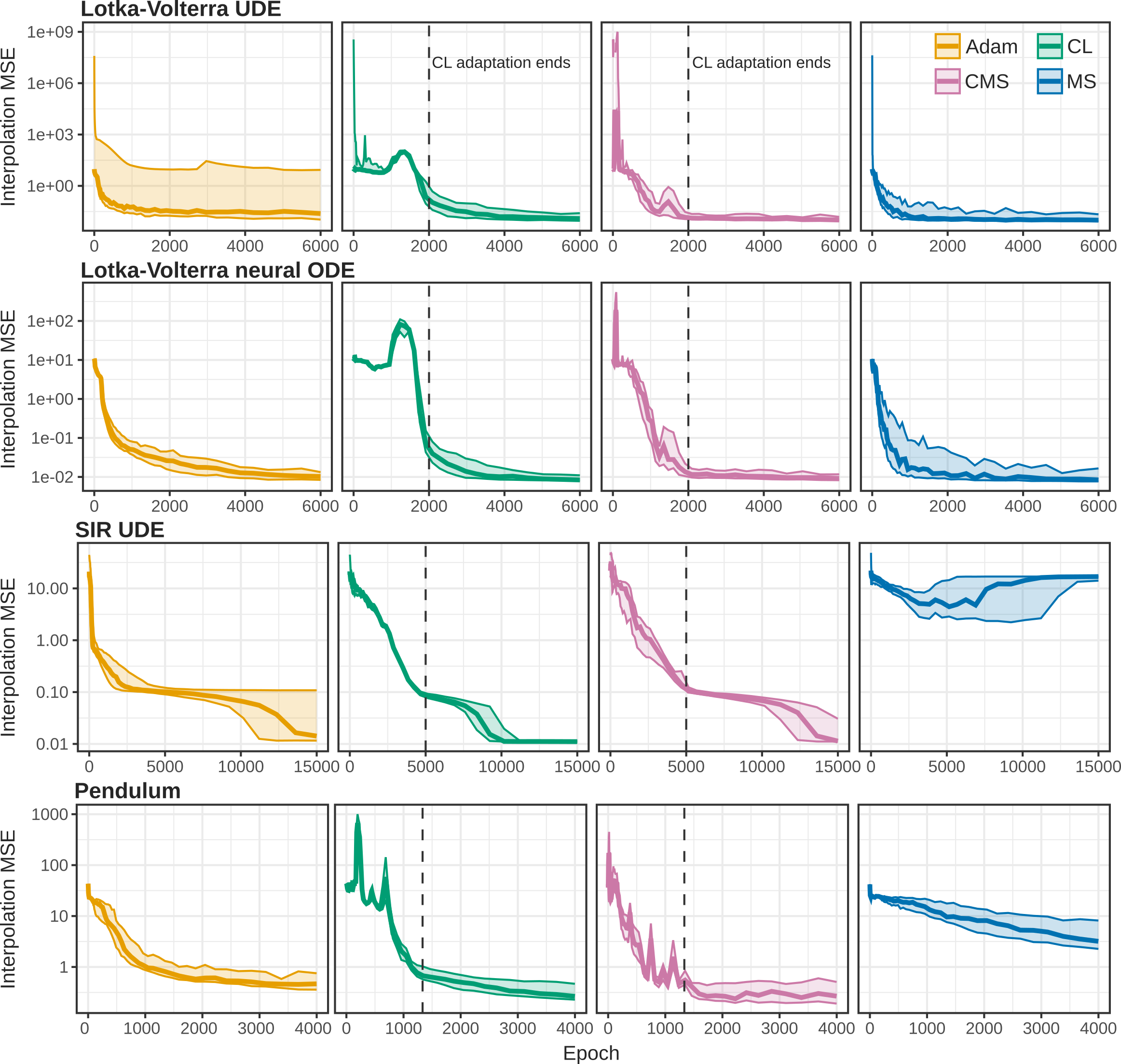}
    \caption{Interpolation-MSE traces over training epochs for the Lotka--Volterra UDE (top), Lotka--Volterra NODE (upper middle), SIR UDE (lower middle), and pendulum UDE (bottom) models. Results are based on 100 multistarts with random parameter initializations shown on a $\log_{10}$ scale. Solid lines denote the median and shaded intervals the 25th--75th percentiles across multistarts. Dashed vertical lines indicate the end of curriculum adaptation for the relevant strategies. Adam+BFGS is excluded because we could not extract training traces from the BFGS \texttt{Fides} optimizer.}
    \label{fig:training_traces}
\end{figure}

As CL performed on par with CMS for the SIR and Lotka--Volterra NODE models (Fig.~\ref{fig:ude_benchmarks}), we next compared their single-shooting (Eq.~\ref{eq:objective}) interpolation-MSE training curves over epochs (Fig.~\ref{fig:training_traces}). CMS generally converged fastest, reaching a low MSE within roughly one-third of the training epochs, whereas CL typically required at least two-thirds. The exception was the SIR model (Fig.~\ref{fig:training_traces}c), where CL reached a low MSE within two-thirds of the epochs, whereas CMS required the full $15\,000$ epochs. SIR was also the only model for which the CL MSE decreased monotonically and, importantly, the model dynamics are non-oscillatory. For the other models, the CL interpolation MSE spiked during early curriculum stages (Fig.~\ref{fig:training_traces}). This suggested that CL can overfit when early curriculum stages are fitted to data subsets that are not representative of the full time-series data, for example when they cover only part of an oscillation period. By contrast, because CMS is trained on the full time series, it can explain why it often converged faster.

\clearpage
\subsection{Additional details on mechanistic benchmarks}
\label{app:mech_benchmarks}

This section provides additional discussion of curriculum learning for the mechanistic ODE benchmarks.

\begin{figure}[ht]
    \centering
    \includegraphics[width=0.99\linewidth]{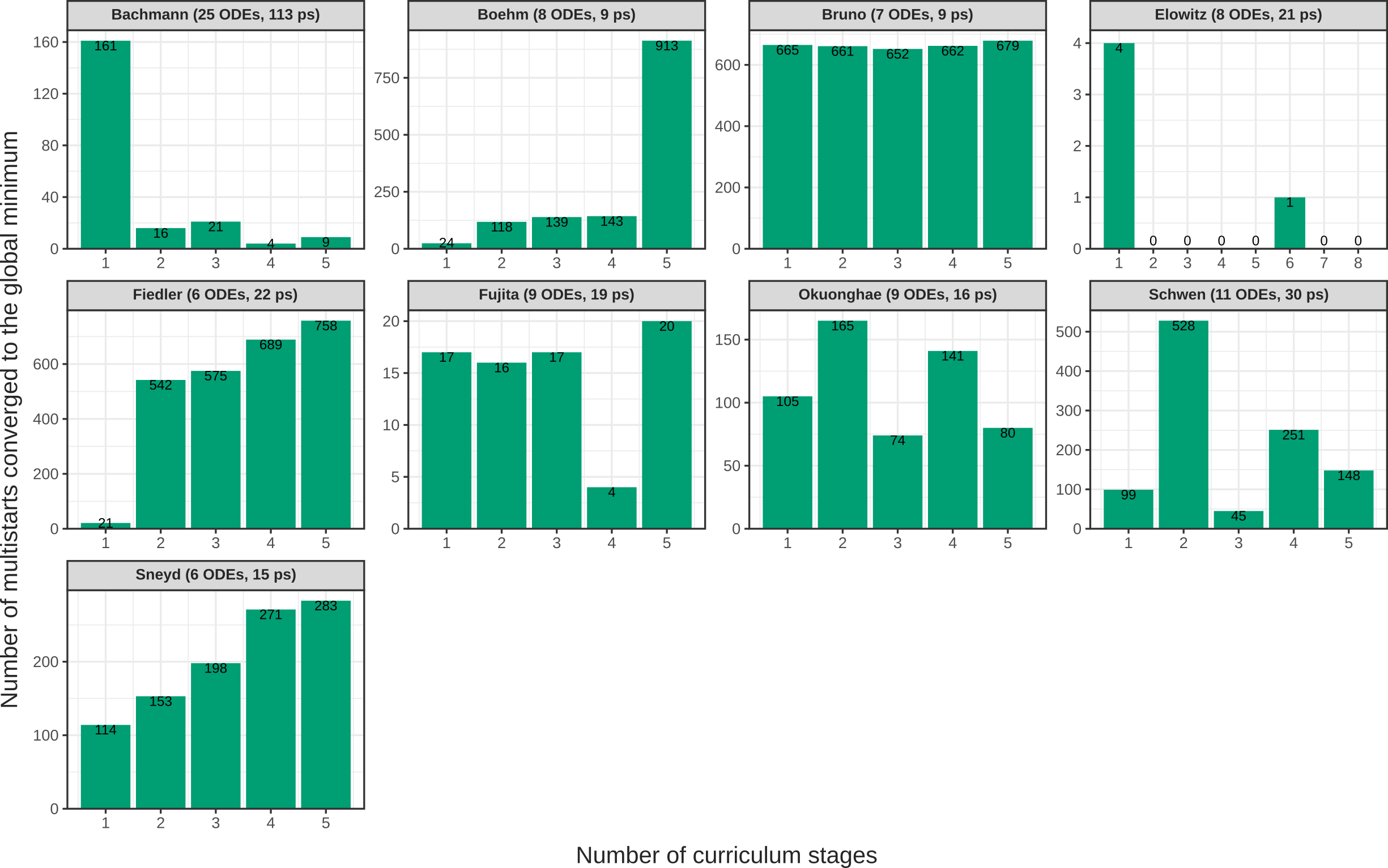}
    \caption{Number of runs converging to the global minimum out of $1000$ multistarts for curriculum learning with different numbers of curriculum stages. For example, for the Boehm model, $24$ runs converge with one curriculum stage, corresponding to standard single-shooting training, whereas $913$ runs converge with five stages. The models are the same as in Fig.~\ref{fig:mech_benchmarks} and Tab.~\ref{tab:mechanistic_models}, with the addition of the Bachmann model. For each model, the information within parenthesis denotes the number of model states and number of parameters (ps) to estimate.}
    \label{fig:cl_mech_extra}
\end{figure}

Since CL is competitive with CMS on a subset of both mechanistic ODE benchmarks and UDE/NODE models (Figs.~\ref{fig:ude_benchmarks}--\ref{fig:mech_benchmarks}), we asked under which conditions CL performs well. Although no single definitive pattern emerges, three models where CL performs strongly (Boehm, Fiedler, and Sneyd; Fig.~\ref{fig:mech_benchmarks}b) share a common feature: each curriculum stage contains all observables (measured model outputs) and all simulation conditions, i.e. experiments in which the model is simulated under different control parameters. This suggests that CL works well in cases where the number of curriculum stages can be chosen such that early curriculum stages are representative of the full problem. To further test this hypothesis, we considered an additional benchmark model, the Bachmann model, which has multiple simulation conditions, many of which appear only at late time points and therefore only in later CL stages. In this case, CL performed poorly (Fig.~\ref{fig:cl_mech_extra}).

While representativeness provides useful intuition for when CL performs well, it does not determine how to chose the number of curriculum stages. Adding more total representative stages improves performance for some mechanistic models (Boehm, Fiedler, and Sneyd), whereas others (Okuonghae and Schwen) perform best with only two stages (Fig.~\ref{fig:cl_mech_extra}). Similarly, for the NODE/UDE benchmarks, using more representative curriculum stages did not consistently improve training performance for 2/4 models (Tab.~\ref{tab:cl_tuning}). These results suggest that CL is most effective when the problem changes gradually across stages, but that the optimal number of stages remains problem-dependent.

In summary, as CL has fewer hyperparameters and is faster than CMS, it remains attractive when it performs well. Our results suggest that this occurs when early curriculum stages are representative of the full time series, for example when all observables are present in early stages or when the dynamics are relatively simple (e.g. non-oscillatory). Otherwise, CMS performs better while remaining competitive on easier problems. A relevant future direction is to develop practical guidance on when CL is sufficient.

\subsection{Complementary UDE/NODE benchmarks in Julia}
\label{sec:julia_ude_benchmark}

To assess the robustness of CMS to the software ecosystem and gradient-computation method, we complemented the main (Sec.~\ref{sec:experiments}) Python UDE/NODE benchmarks with experiments in the Julia SciML ecosystem~\citep{rackauckas_universal_2021}, using \texttt{PEtab.jl}~\citep{persson_petabjl_2025}. As described in App.~\ref{app:experiment_details}, with \texttt{PEtab.jl} we used forward-mode automatic differentiation via \texttt{ForwardDiff.jl}~\citep{revels_forward-mode_2016} to compute gradients. In this setting, sensitivities are computed alongside the discretized ODE solution, so ODE solver error control also acts on the dual components carrying gradient information. This can therefore be considered an optimize-then-discretize gradient-computation scheme. This contrasts with the main Python \texttt{diffrax}-based benchmarks in Sec.~\ref{sec:experiments} (Figs.~\ref{fig:lv_ude}--\ref{fig:ude_benchmarks}), where gradients were computed using \texttt{RecursiveCheckpointAdjoint}, corresponding to a discretize-then-optimize scheme.

To avoid the computational cost of re-tuning all training strategies, we used the same hyperparameters as in Sec.~\ref{sec:experiments} (Tabs.~\ref{tab:adam_tuning}--\ref{tab:cl_ms_tuning}) for single-shooting Adam, curriculum learning (CL), multiple shooting (MS), and curriculum multiple shooting (CMS). We did not consider the Adam+BFGS training strategy. As in the main experiments, we ran $100$ multistarts. We considered the Lotka--Volterra UDE (Eq.~\ref{eq:lv_ude_eq}), Lotka--Volterra NODE (Eq.~\ref{eq:lv_node_eq}), and SIR UDE (Eq.~\ref{eq:sir_eq}). We also attempted the double-pendulum UDE (Eq.~\ref{eq:pendulum_ude}), but excluded it because a single multistart required more than five days and exceeded our compute budget. This substantial runtime is because forward-mode automatic differentiation scales poorly with the number of estimated parameters~\citep{sapienza_differentiable_2025}, and the double pendulum has the largest neural network. All models were trained for the same number of epochs as in the main experiments (App.~\ref{app:experiment_details}). Noticeably, one implementation detail differs from the main CMS experiments. In the main text, we penalize the full overlap between adjacent windows. In the \texttt{PEtab.jl} implementation, CMS is implemented at the abstraction level of the PEtab parameter-estimation standard~\citep{schmiester_petabinteroperable_2021}, which only allowed us to penalize the first overlapping point, i.e. the MS window boundaries.

\begin{figure}[ht]
    \centering
    \includegraphics[width=0.99\linewidth]{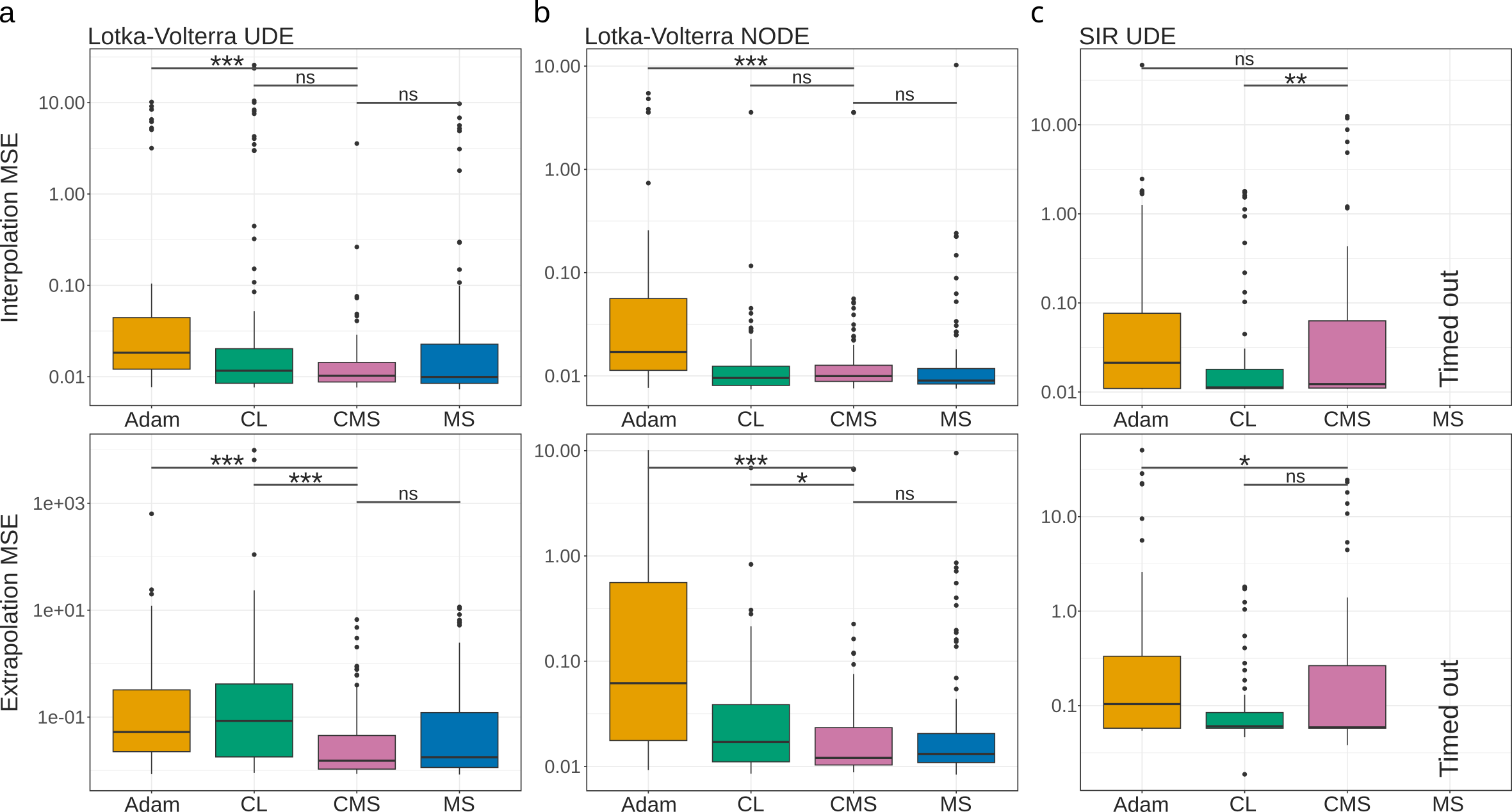}
    \caption{Complementary UDE/NODE benchmark results from $100$ multistarts with random parameter initializations using the Julia SciML ecosystem. Interpolation MSE (top) and extrapolation MSE (bottom) are shown on a $\log_{10}$ scale for the (a) Lotka--Volterra UDE, (b) Lotka--Volterra NODE, and (c) SIR UDE models. P-values were computed using the Wilcoxon rank-sum test and Holm-corrected within each model. Significance levels are denoted by ns ($p \geq 0.05$), * ($p < 0.05$), ** ($p < 0.01$), and *** ($p < 0.001$).}
    \label{fig:julia_benchmark}
\end{figure}

As in the main experiments (Sec.~\ref{sec:experiments}), CMS was consistently top-performing in terms of median extrapolation MSE (Fig.~\ref{fig:julia_benchmark} and Tab.~\ref{tab:julia_ude_res}). For the Lotka--Volterra UDE (Fig.~\ref{fig:julia_benchmark}a), CMS performed similarly to the second-best method, multiple shooting (0.0152 vs.\ 0.0176; $p = 0.44$), and significantly outperformed CL and Adam in extrapolation. For the Lotka--Volterra NODE (Fig.~\ref{fig:julia_benchmark}b), CMS again matched the second best MS strategy (0.0121 vs.\ 0.0131; $p = 0.56$). We note that plain single-shooting Adam differed in performance from the main Python benchmark, with interpolation MSE increasing from $0.0103$ to $0.0171$ and extrapolation MSE from $0.016$ to $0.062$ (Tabs.~\ref{tab:lv_node_res} and~\ref{tab:julia_ude_res}). This highlights that the software ecosystem and gradient-computation method can affect training even when using the same hyperparameters and ODE-solver algorithm. Finally, for the SIR UDE (Fig.~\ref{fig:julia_benchmark}c), CMS matched CL in extrapolation MSE, which was the best-performing strategy (0.0589 vs.\ 0.0589; $p = 0.69$), while having slightly worse interpolation MSE (0.0123 vs.\ 0.0113; $p = \num{1.6e-3}$). The MS strategy timed out for the SIR UDE, as no multistart completed within the 48-hour compute budget. This reflects poor MS training on this model, which inspection of training traces confirmed.

In summary, CMS remains robust when changing both the software ecosystem and gradient-computation method. This robustness holds despite using a different continuity-penalty scheme, where re-tuning could potentially improve performance. The differences in plain Adam performance between the Julia and Python benchmarks further show that implementation details can affect training performance and the optimal hyperparameter configuration.

\begin{table}[ht]
    \centering
    \caption{Benchmark results for the Lotka--Volterra UDE, Lotka--Volterra NODE, and SIR UDE models in the Julia SciML ecosystem benchmark. Interpolation and extrapolation MSE are reported as $\mathrm{median} \pm \mathrm{median\ absolute\ deviation}$ for each training strategy. The best interpolation and extrapolation values within each model are shown in bold.}
    \label{tab:julia_ude_res}
    \begin{tabular}{llcc}
        \toprule
        Model & Strategy & Interpolation MSE & Extrapolation MSE \\
        \midrule
        Lotka--Volterra UDE
        & Adam  & $0.0183 \pm 0.0122$ & $0.0525 \pm 0.0611$ \\
        & CL    & $0.0116 \pm 0.00504$ & $0.0851 \pm 0.111$ \\
        & CMS & $0.0102 \pm 0.00260$ & $\mathbf{0.0152 \pm 0.00805}$ \\
        & MS    & $\mathbf{0.00995 \pm 0.00303}$ & $0.0176 \pm 0.0123$ \\
        \midrule
        Lotka--Volterra NODE
        & Adam  & $0.0171 \pm 0.0116$ & $0.0619 \pm 0.0754$ \\
        & CL    & $0.00954 \pm 0.00243$ & $0.0171 \pm 0.0107$ \\
        & CMS & $0.00994 \pm 0.00199$ & $\mathbf{0.0121 \pm 0.00361}$ \\
        & MS    & $\mathbf{0.00901 \pm 0.00131}$ & $0.0131 \pm 0.00559$ \\
        \midrule
        SIR UDE
        & Adam  & $0.0213 \pm 0.0156$ & $0.104 \pm 0.0700$ \\
        & CL    & $\mathbf{0.0113 \pm 0.000667}$ & $0.0605 \pm 0.00563$ \\
        & CMS & $0.0123 \pm 0.00202$ & $\mathbf{0.0589 \pm 0.0169}$ \\
        \bottomrule
    \end{tabular}
\end{table}

\clearpage
\setcounter{figure}{0}
\setcounter{table}{0}
\setcounter{equation}{0}

\section{Benchmark models and data}
\label{app:benchmark_details}

\setcounter{figure}{0}

This section provides additional details for each benchmark model and its associated time-series dataset.

\subsection{Lotka--Volterra System}

The Lotka--Volterra system is a simplified model of predator ($x$) and prey ($y$) population dynamics given by

\begin{equation}
    \label{eq:lv_eq}
    \begin{split}
        \frac{\mathrm{d}x}{\mathrm{d}t} &= \alpha x - \beta x y, \\
        \frac{\mathrm{d}y}{\mathrm{d}t} &= -\delta y + \gamma x y,
    \end{split}
\end{equation}

where $\alpha$ is the prey birth rate, $\delta$ is the predator death rate, and ($\beta, \gamma$) describe the interaction strength.

\textbf{Model architectures}. We consider two model classes: a UDE and a black-box NODE. Following~\citet{rackauckas_universal_2021}, for the UDE we use a structure that incorporates known birth and death terms, while learning the interaction terms:

\begin{equation}
    \label{eq:lv_ude_eq}
    \begin{split}
        \frac{\mathrm{d}x}{\mathrm{d}t} &= \alpha x(t) + \mathbf{NN}_1\big(x(t), y(t); \pmb{\theta}_n\big), \\
        \frac{\mathrm{d}y}{\mathrm{d}t} &= -\delta y(t) + \mathbf{NN}_2\big(x(t), y(t); \pmb{\theta}_n\big).
    \end{split}
\end{equation}

For the black-box NODE, the full right-hand side (RHS) is represented by the neural network:

\begin{equation}
    \label{eq:lv_node_eq}
    \begin{split}
        \frac{\mathrm{d}x}{\mathrm{d}t} &= \mathbf{NN}_1\big(x(t), y(t); \pmb{\theta}_n\big), \\
        \frac{\mathrm{d}y}{\mathrm{d}t} &= \mathbf{NN}_2\big(x(t), y(t); \pmb{\theta}_n\big).
    \end{split}
\end{equation}

For both models, $\mathbf{NN}(x, y; \pmb{\theta}_n)$ is a feed-forward neural network that takes $(x,y)$ as input and has layer widths $[2,10,10,2]$. Each layer includes a bias term; swish activations are used in all hidden layers, and the identity function is used in the output layer.

\begin{figure}[ht]
    \centering
    \includegraphics[width=0.80\linewidth]{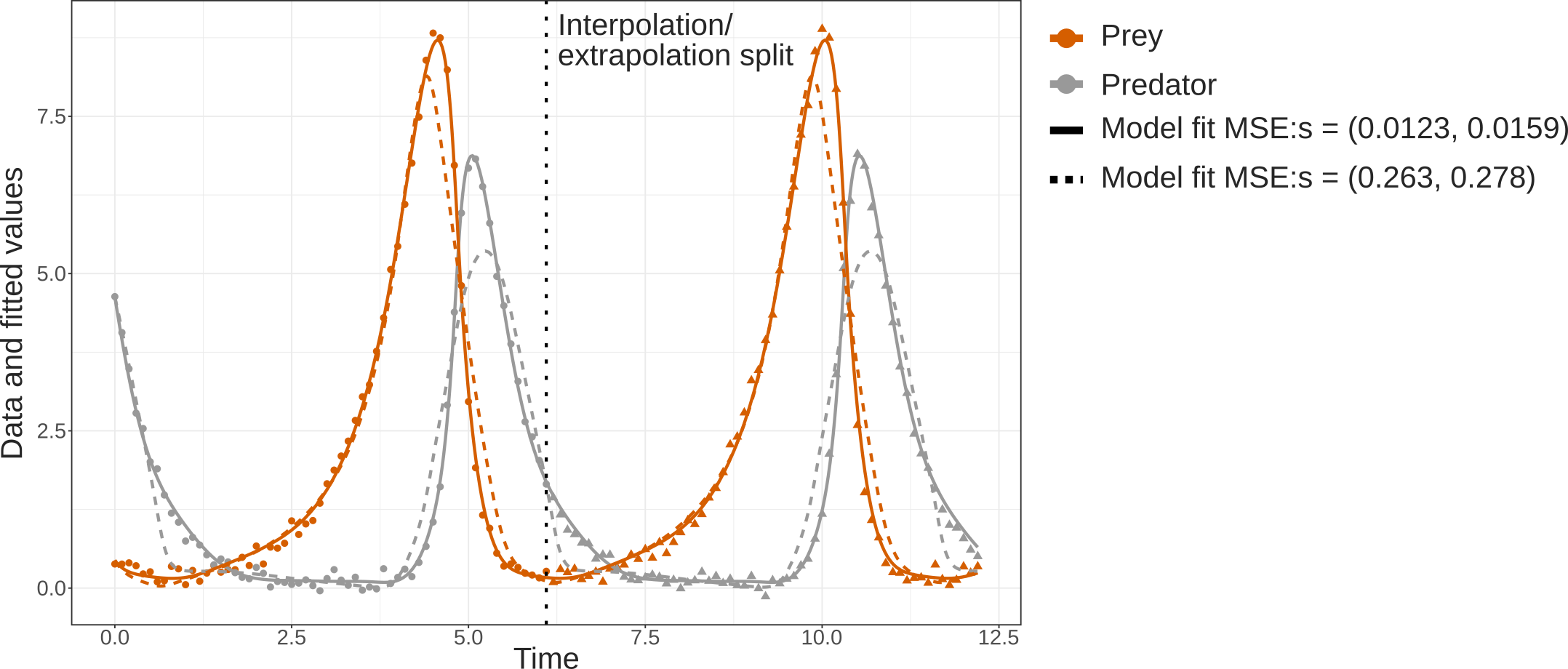}
    \caption{Lotka--Volterra UDE and NODE training and validation data for the prey and predator observables. Lines show two representative model fits with different interpolation/training and extrapolation/validation errors. MSE values are reported in the legend as (interpolation, extrapolation).}
    \label{fig:lv_data}
\end{figure}

For training, we follow~\citep{ko_homotopy-based_2023} and assume known initial values $(x_0, y_0) = (0.44249296, 4.6280594)$. For the UDE, we estimate the mechanistic parameters $(\alpha, \delta)$ and the neural-network parameters $\pmb{\theta}_n$, while for the NODE we estimate only $\pmb{\theta}_n$.

\textbf{Training and validation data}. For both Lotka--Volterra models, we use simulated data. Following~\citep{ko_homotopy-based_2023}, for the main experiments in Figs.~\ref{fig:lv_ude} and \ref{fig:ude_benchmarks}, we simulate Eq.~\ref{eq:lv_eq} on the interval $[0.0, 12.2]$ with sampling interval $\Delta t = 0.1$, initial values $(x_0, y_0) = (0.44249296, 4.6280594)$, and parameter values $(\alpha, \beta, \gamma, \delta) = (1.3, 0.9, 0.8, 1.8)$. We noise-corrupt the measurements with Gaussian noise with zero mean and standard deviation equal to $5\%$ of the mean of each trajectory. The data are split into training and validation using a 50/50, so we train on measurements in $t \in [0.0, 6.1]$ ($62 \times 2$ measurements) and validate on measurements in $t \in [6.2, 12.2]$ ($62 \times 2$ measurements). The data, together with representative fits for different MSE values, are shown in Fig.~\ref{fig:lv_data}.

For the noise-robustness experiments (Fig.~\ref{fig:lv_ude_noise}), we generate data as above but with noise levels of $10\%$, $20\%$, and $50\%$. For the simulation-horizon experiments, we instead simulate the system on $[0.0, 18.2]$ and $[0.0, 22.2]$, using the same sampling interval $\Delta t = 0.1$. For both robustness experiments, we apply a 50/50 train-validation split.

\subsection{SIR UK UDE System}

The SIR (susceptible, infected, recovered) UDE model was introduced in~\citep{dandekar_machine_2020}, where Dandekar et al.\ augmented a standard SIR model to better capture the early dynamics of the COVID-19 outbreak.

\textbf{Model architecture}. We use essentially the same model structure as in~\citep{dandekar_machine_2020}, with dynamics

\begin{equation}
    \label{eq:sir_eq}
    \begin{split}
        \frac{\mathrm{d}S}{\mathrm{d}t} &= - \frac{\beta S(t) I(t)}{N}, \\
        \frac{\mathrm{d}I}{\mathrm{d}t} &= \frac{\beta S(t) I(t)}{N} - \gamma I(t) - \mathbf{NN}_1\big(S(t)/N, I(t)/N, R(t)/N; \pmb{\theta}_n\big) \cdot I(t), \\
        \frac{\mathrm{d}R}{\mathrm{d}t} &= \gamma I(t) + \delta T(t), \\
        \frac{\mathrm{d}T}{\mathrm{d}t} &= \mathbf{NN}_1\big(S(t)/N, I(t)/N, R(t)/N; \pmb{\theta}_n\big) \cdot I(t) - \delta T(t),
    \end{split}
\end{equation}

where $T$ denotes the quarantined population, $(\beta, \gamma, \delta)$ are unknown mechanistic parameters, $N = S + I + R$ is the total population, and $\mathbf{NN}$ is a neural network parameterized by $\bm{\theta}_n$ that represents a time-varying quarantine strength. Following~\citep{dandekar_machine_2020}, $\mathbf{NN}$ is a feed-forward network with layer widths $[3,10,1]$, bias terms in each layer, swish activations in the hidden layers, and the identity function in the output layer. Our implementation differs from~\citet{dandekar_machine_2020} in one respect: we use normalized inputs $(S/N, I/N, R/N)$ rather than the unnormalized states. This choice is motivated by the observation that unnormalized inputs are highly multi-scale (e.g. $S \propto 10^7$ while $R \propto 10^3$), leading to large gradients only in the first layer and vanishing gradients in subsequent layers.

For model training, we use data from the United Kingdom (UK) (see below) and take the initial time to be the day on which approximately 500 infected cases were observed, corresponding to initial values $(S, I, R, T) = (66000000, 762, 26, 10)$. Because the system is only partially observed, with no direct measurements of $S$ or $T$, we match model outputs to the observables

\begin{equation}
    \label{eq:sir_obs}
    \begin{split}
        obs_1(t) &= I(t) + T(t), \\
        obs_2(t) &= R(t).
    \end{split}
\end{equation}

We estimate the mechanistic parameters $(\beta, \gamma, \delta)$ and the neural-network parameters $\bm{\theta}_n$.

\begin{figure}[ht]
    \centering
    \includegraphics[width=0.80\linewidth]{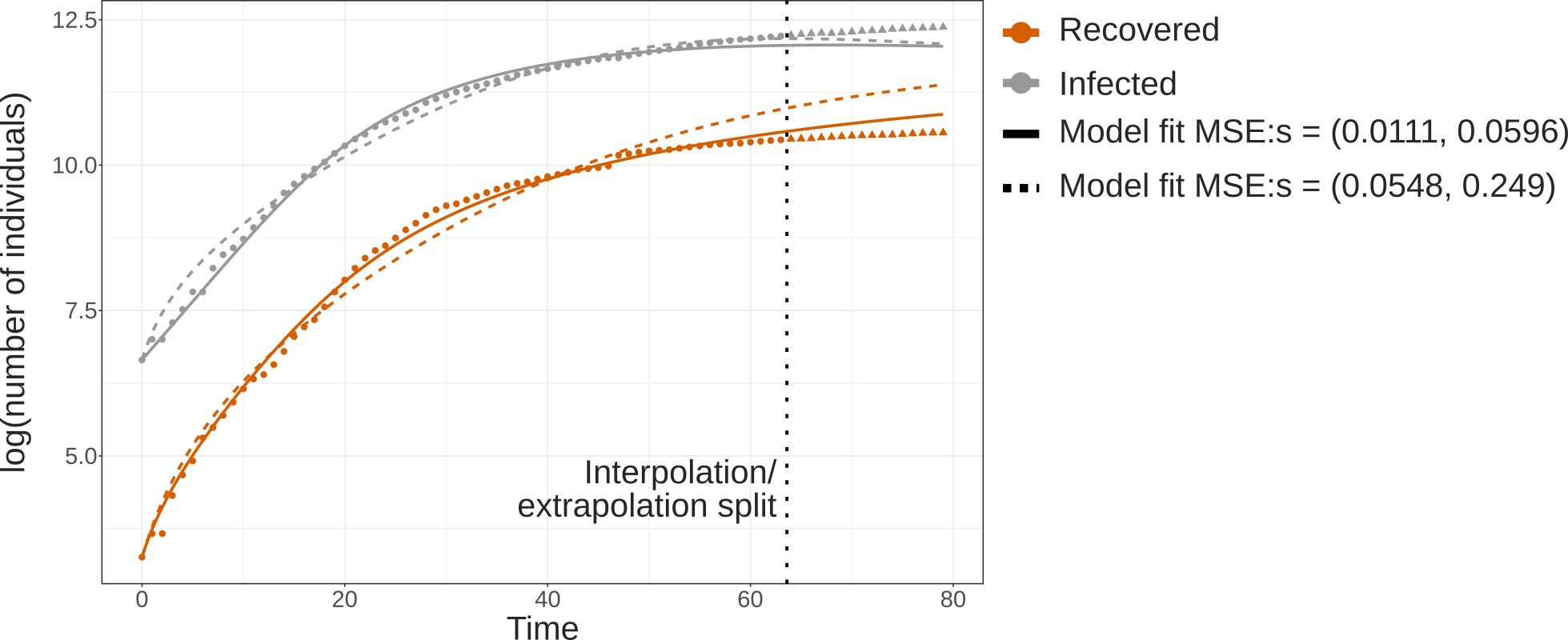}
    \caption{SIR UDE training and validation data for the infected and recovered observables (Eq.~\ref{eq:sir_obs}). Lines show two representative model fits with different interpolation/training and extrapolation/validation errors. MSE values are reported in the legend as (interpolation, extrapolation).}
    \label{fig:sir_data}
\end{figure}

\textbf{Training and validation data}. We use the same UK dataset as in~\citep{dandekar_machine_2020}, originally obtained from the CSSE at Johns Hopkins University. The dataset contains 79 daily measurements of infected cases ($I_{\mathrm{data}}$) and recovered cases ($R_{\mathrm{data}}$). As in~\citep{dandekar_machine_2020}, we use a MSE loss comparing measurements on log-scale:

\begin{equation}
    \label{eq:sir_loss}
    \mathcal{L}
    =
    \frac{1}{N_{\mathrm{obs}}}
    \sum_j
    \left(
        \big(\log(obs_1(t_j)) - \log(I_{\mathrm{data}}(t_j))\big)^2
        +
        \big(\log(obs_2(t_j)) - \log(R_{\mathrm{data}}(t_j))\big)^2
    \right),
\end{equation}

where $N_{\mathrm{obs}}$ denotes the total number of observed measurements across both observables. Note, the $I_{data}$ is assumed to include both the infected ($I$) and quarantine ($T$) population from the model. We use a 70/30 split, thus training on measurements in $t \in [0,63]$ ($64 \times 2$ measurements) and validate on measurements in $t \in [64,79]$ ($32 \times 2$ measurements), with sampling interval $\Delta t = 1.0$. The data with fits for different MSE values are shown in Fig.~\ref{fig:sir_data}.

\subsection{Double Pendulum System}

The double pendulum is a canonical system in classical mechanics. It is governed by four states $(\theta_1,\theta_2,\omega_1,\omega_2)$, describing the angles and angular velocities of the two pendulums relative to the vertical. Its dynamics are naturally second order, but can be written in first-order form as

\begin{equation}
    \label{eq:pendulum_eq}
    \begin{split}
        \frac{\mathrm{d}\theta_i}{\mathrm{d}t} &= \omega_i(t), \qquad i=1,2, \\
        \frac{\mathrm{d}\omega_i}{\mathrm{d}t} &= f_i\big(\omega_1(t),\omega_2(t),\theta_1(t),\theta_2(t)\big), \qquad i=1,2,
    \end{split}
\end{equation}

where $f_1$ and $f_2$ can be derived from the Lagrangian formulation of classical mechanics; the full expressions are given in~\citep{ko_homotopy-based_2023}. The double pendulum is a challenging oscillatory system to learn and therefore provides a strong stress test for our CMS method.

\textbf{Model architecture}. Following~\citep{ko_homotopy-based_2023}, we use a physics-informed UDE structure of the form

\begin{equation}
    \label{eq:pendulum_ude}
    \begin{split}
        \frac{\mathrm{d}\theta_i}{\mathrm{d}t} &= \omega_i, \qquad i=1,2, \\
        \frac{\mathrm{d}\omega_i}{\mathrm{d}t} &= \mathbf{NN}_i\big(\omega_1(t),\omega_2(t),\theta_1(t),\theta_2(t);\bm{\theta}_n\big), \qquad i=1,2,
    \end{split}
\end{equation}

where $\mathbf{NN}$ is a feed-forward network with layer widths $[4,50,50,2]$, bias terms in each layer, swish activations in the hidden layers, and the identity function in the output layer. The data are fully observed, so we use initial values $(\theta_1,\theta_2,\omega_1,\omega_2)=(-1.13,-1.30,-0.53,-6.15)$. For training, we estimate the neural-network parameters $\bm{\theta}_n$.

\begin{figure}[ht]
    \centering
    \includegraphics[width=0.95\linewidth]{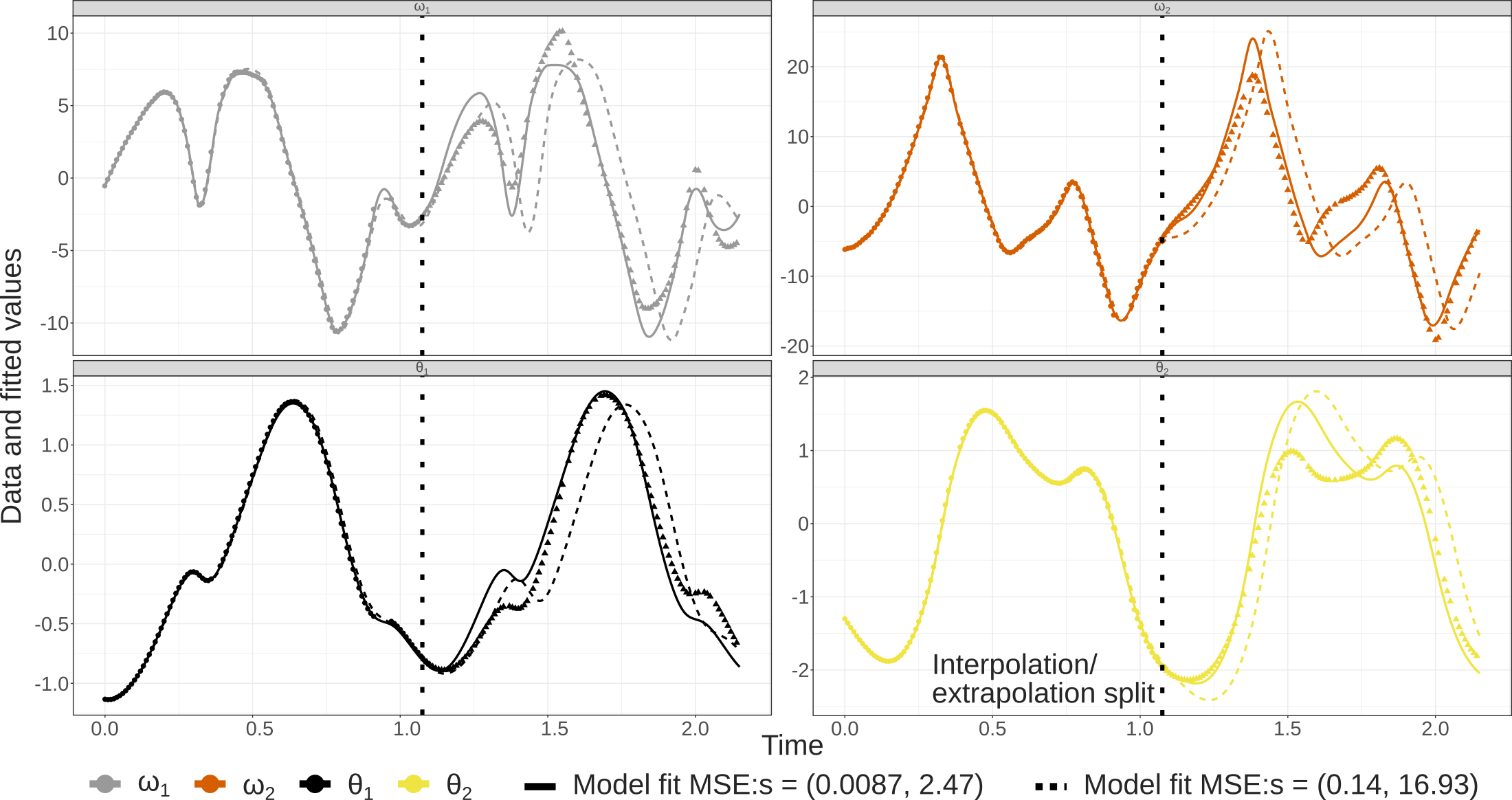}
    \caption{Pendulum training and validation data for the $(\omega_1, \omega_2, \theta_1, \theta_2)$ observables. Lines show two representative model fits with different interpolation/training and extrapolation/validation errors. MSE values are reported in the legend as (interpolation, extrapolation).}
    \label{fig:pendulum_data}
\end{figure}

\textbf{Training and validation data}. We use the same real double-pendulum dataset as in~\citet{ko_homotopy-based_2023}, originally generated by~\citet{schmidt_distilling_2009}. While the double pendulum can exhibit chaotic behaviour, we follow~\citet{ko_homotopy-based_2023} and evaluate on the same short trajectory segment. This short-horizon setting limits the extent to which long-term chaotic divergence affects the comparison of training strategies. We use a 70/30 split, training on measurements in $t\in[0.0,1.71]$ ($160 \times 4$ measurements) and validating on measurements in $t\in[1.72,2.14]$ ($40 \times 4$ measurements). The data with representative fits are shown in Fig.~\ref{fig:pendulum_data}.

\subsection{Mechanistic ODE Models}

The mechanistic ODE models are drawn from the PEtab benchmark collection, which contains a diverse set of models with real time-series data~\citep{hass_benchmark_2019, schmiester_petabinteroperable_2021}. The collection primarily comprises models from cell biology (e.g. cell signalling and gene regulation) and epidemiology (e.g. COVID-19 progression). From a numerical perspective, these models present a range of challenging training features. Most exhibit stiff dynamics arising from disparate time scales (e.g. fast and slow chemical reactions)~\citep{persson_petabjl_2025}, state variables spanning multiple orders of magnitude, and non-identifiability leading to sloppy parameter landscapes~\citep{hass_benchmark_2019, gutenkunst_universally_2007}. Most models are also only partially observed, with observables typically given as functions of multiple state variables (e.g. $obs_1 = x + y$). Measurements are noisy and often irregularly sampled, with different observables measured at different time points. A subset of models also includes multiple experimental conditions, requiring simulations under different control parameter values for different measurements. We selected eight representative models of varying size from this collection, whose key properties are summarized in Tab.~\ref{tab:mechanistic_models}. Overall, they provide a robust stress test for our training strategy.

\begin{table}[ht]
    \centering
    \caption{Summary of the eight mechanistic ODE benchmark models fitted to real data. For each model, the number of states, parameters to estimate (Params.), measurements (Meas.), observables (Obs.), experimental conditions (Cond.), as well as the application domain is reported.}
    \label{tab:mechanistic_models}
    \begin{tabular}{lcccccc}
        \toprule
        Model & States & Params. & Meas. & Obs. & Cond. & Domain \\
        \midrule
        Boehm     & 8  & 9  & 48  & 3 & 1  & Biology \\
        Bruno     & 7  & 13 & 77  & 5 & 6  & Biology \\
        Elowitz   & 8  & 21 & 58  & 1 & 1  & Biology \\
        Fiedler   & 6  & 22 & 72  & 2 & 3  & Biology \\
        Fujita    & 9  & 19 & 144 & 3 & 6  & Biology \\
        Okuonghae & 2  & 9  & 92  & 1 & 1  & Epidemiology \\
        Schwen    & 11 & 30 & 286 & 4 & 19 & Biology \\
        Sneyd     & 6  & 15 & 135 & 1 & 9  & Biology \\
        \bottomrule
    \end{tabular}
\end{table}

\clearpage
\setcounter{figure}{0}
\setcounter{table}{0}
\setcounter{equation}{0}

\section{Additional experiment details}
\label{app:experiment_details}

This section provides additional experimental details, including the software and ODE solvers used, parameter initialization procedures, implementation details for the training strategies evaluated in Sec.~\ref{sec:experiments}, and the hyperparameter optimization and selection procedures.

\subsection{Software implementation}

All training strategies considered in this study were implemented by us. The main NODE and UDE benchmarks were implemented in Python using \texttt{diffrax} and \texttt{equinox}~\citep{kidger_neural_2022}. Julia was used for the mechanistic ODE benchmarks, as well as for complementary robustness experiments on the UDE benchmarks, using \texttt{PEtab.jl}~\citep{persson_petabjl_2025}. Code for all benchmarks is provided in the attached zip file and will be made available in a public repository upon de-anonymization.

For the Python \texttt{diffrax}-based benchmarks, we used the explicit \texttt{Tsit5} ODE solver~\citep{tsitouras_rungekutta_2011}, with tolerances $(\mathrm{atol}, \mathrm{rtol}) = (10^{-8}, 10^{-8})$. Gradients were computed using \texttt{RecursiveCheckpointAdjoint}, corresponding to a discretize-then-optimize approach in which gradients are obtained by differentiating through the numerical solver.

For the Julia \texttt{PEtab.jl}-based benchmarks, we used \texttt{Tsit5} with $(\mathrm{atol}, \mathrm{rtol}) = (10^{-8}, 10^{-8})$ for the UDE robustness experiments, and the stiff \texttt{Rodas5P} solver~\citep{steinebach_construction_2023}, also with $(\mathrm{atol}, \mathrm{rtol}) = (10^{-8}, 10^{-8})$, for the mechanistic ODE benchmarks. This choice was motivated by previous results showing that models from the PEtab benchmark collection are generally stiff~\citep{persson_petabjl_2025}. For gradient computation, we used forward-mode automatic differentiation via \texttt{ForwardDiff.jl}~\citep{revels_forward-mode_2016}, where derivatives are propagated through the numerical solve using dual numbers. In this setting, sensitivities are computed alongside the discretized ODE solution, so solver error control also acts on the dual components carrying gradient information. This has been argued to yield more reliable gradients~\citep{sapienza_differentiable_2025}, although other work suggests that discretize-then-optimize approaches can produce more accurate gradients~\citep{onken_discretize-optimize_2020}.

\subsection{Computational resources}
\label{app:computational_resources}

All benchmarks were run on CPUs on an institutional cluster. The average time for a Lotka-Volterra training run was between 3 and 15 minutes depending on the training approach, and the requested amount of memory was 16GB.  The SIR model trained in under 15 minutes on average also with a requested memory of 16GB.  The double pendulum model trained in under 10 minutes on average, with the exception of the Adam+BFGS training approach, where some runs timed out completely at a cut-off of 2 hours. The requested memory for pendulum training runs was 32GB.

\subsection{Initialization of estimated parameters}

For both the mechanistic and NODE/UDE models, we performed multiple training runs from randomly sampled initial parameter values for all training strategies. For the mechanistic models, parameters were initialized using Latin hypercube sampling, which has been shown to outperform uniform random sampling in this setting~\citep{raue_lessons_2013}. All mechanistic models provide parameter bounds, and sampling was performed within those bounds.

For the UDE models with unknown mechanistic parameters, namely the SIR and Lotka--Volterra UDEs, mechanistic parameters were Latin hypercube sampled on the $\log_{10}$ scale within two orders of magnitude of a reference parameter vector $\bm{\theta}_{\mathrm{ref}}$. For the Lotka--Volterra models, $\bm{\theta}_{\mathrm{ref}}$ was the true parameter vector; for the SIR model, it was the initial parameter vector used for training in~\citep{dandekar_machine_2020}. Neural-network parameters were initialized with zero biases and Glorot-uniform weights~\citep{glorot_understanding_2010}, using a gain of $0.1$. This gain is smaller than the default, as NODEs often benefit from initialization at smaller parameter values~\citep{kidger_neural_2022}.

\subsection{Training budgets}

\textbf{NODE and UDE benchmarks.}
Following~\citet{ko_homotopy-based_2023}, we trained the Lotka--Volterra UDE and NODE models for $6000$ epochs, and the double-pendulum UDE for $4000$ epochs. We trained the SIR UDE for $15\,000$ epochs. This budget was chosen to allow the training strategies to converge, while remaining substantially smaller than the $60\,000$ epochs used in the original publication~\citep{dandekar_machine_2020}. To ensure a fair comparison, all training strategies were run for the same number of epochs for each model.

\textbf{Mechanistic ODE benchmarks.}
For the mechanistic ODE benchmarks, we used a Newton trust-region optimizer with adaptive step lengths and automatic termination criteria. Therefore, different methods did not necessarily run for the same number of iterations. As is standard for these benchmarks~\citep{persson_petabjl_2025}, we allowed a maximum of $1000$ optimizer iterations per run.

\subsection{Training strategy implementation details}

This section provides additional implementation details for the training strategies evaluated in Sec.~\ref{sec:experiments}, as well as the output regularization used in the UDE benchmarks.

\subsubsection{Output regularization}

Neural network output regularization was introduced by \citet{philipps_non-negative_2024} to prevent the neural network component of a UDE from dominating the dynamics and effectively overriding the mechanistic component, for example by driving mechanistic parameters toward zero. The regularization term is defined in Eq~.\ref{eq:output_reg}, which is repeated here for completness:

\begin{equation*}
    \lambda_O \left( \int_{t_0}^T \lVert \mathbf{NN}(\mathbf{u}; \bm{\theta}) \rVert_2 \, \mathrm{d}t \right)^2,
\end{equation*}

where $[t_0, T]$ is the simulation interval, $\lVert\cdot\rVert_2$ is the $\ell_2$ norm, $\bm{\theta}$ denotes the neural network parameters, and $\lambda_O$ is a hyperparameter controlling the strength of the regularization. This term is added to the loss function in Eq.~\ref{eq:objective}. In practice, the integral is computed by augmenting the UDE with an additional state $\phi(t)$,

\begin{equation*}
    \frac{\mathrm{d}\phi}{\mathrm{d}t} = \lVert \mathbf{NN}(\mathbf{u}; \bm{\theta}) \rVert_2, \qquad \phi(t_0) = 0,
\end{equation*}

and evaluating $\lambda_O \phi(T)^2$ at the final time.

\subsubsection{Curriculum learning implementation details}

Our curriculum learning strategy is described in Sec.~\ref{sec:curriculum_learning}. It introduces three tunable parameters: the number of curriculum stages $n_c$, the number of epochs per stage, and the stage boundaries $t_i$. Among these, we treated only the number of stages $n_c$ as a tunable hyperparameter; its tuning procedure in App.~\ref{sec:tuning}.

For the other two parameters, we used default settings that performed well across benchmarks. For the number of epochs per stage, given a fixed total training budget, we found it effective to allocate equal numbers of epochs to stages $i=1,\ldots,n_c-1$ over the first third of training, followed by training on the full time series for the remaining two thirds. For the stage boundaries $t_i$, we chose them so that each stage contained approximately the same number of measurements, with at most a one-measurement difference between stages. As discussed in Sec.~\ref{sec:experiments}, the boundaries should also be chosen so that the optimization does not change too abruptly between stages, for which distributing measurements approximately evenly is a reasonable heuristic.

\subsubsection{Multiple shooting details}

Our multiple shooting strategy is described in Sec.~\ref{sec:multiple_shooting} and closely follows the implementation in the \texttt{DiffEqFlux.jl} Julia package~\citep{rackauckas_diffeqfluxjl_2019}. Other variants are possible~\citep{massaroli_differentiable_2021, turan_multiple_2022}, such as treating and solving the multiple shooting objective as an equality-constrained optimization problem~\citep{turan_multiple_2022}. We focus on the approach of~\citet{rackauckas_diffeqfluxjl_2019}, as this is the variant extended by our curriculum multiple shooting method. Multiple shooting introduces three tuning parameters: the number of windows $n_w$, the continuity penalty strength $\lambda_{\mathrm{cont}}$, and the window boundaries $t_i$. Among these, we treat the first two as tunable hyperparameters; their tuning procedure in App.~\ref{sec:tuning}.

For the window boundaries, we followed the same principle as for curriculum learning and chose them so that each window contained approximately the same number of measurements. Importantly, we placed splits only at times $t_i$ for which measurements were available. To ensure that each window over $[t_{i-1}, t_i]$ included all measurements in that interval, we duplicated measurements at the boundaries when needed. It is also possible to split at times without measurements, but for a window to be informative it should contain at least one measurement.

\subsubsection{Curriculum multiple shooting details}

Our curriculum multiple shooting approach is described in detail in Sec.~\ref{sec:cl_ms}. Here, for completeness, we note that we set the number of epochs per stage as in curriculum learning and the initial multiple-shooting window boundaries $t_i$ as in multiple shooting.

\subsubsection{Adam + BFGS details}

For the Adam+BFGS strategy, we first trained the model with Adam and then refined the resulting parameters using the \texttt{Fides} Newton trust-region optimizer with a BFGS Hessian approximation~\citep{frohlich_fides_2022}. For example, for the Lotka--Volterra UDE benchmark (Fig.~\ref{fig:lv_ude}), each Adam-trained run was subsequently optimized with \texttt{Fides}. We chose \texttt{Fides} because it has been shown to perform well for mechanistic ODE models~\citep{frohlich_fides_2022, persson_petabjl_2025}. As a trust-region method, it adapts the step size based on the local accuracy of the trust-region approximation.

\subsubsection{Mechanistic ODE model details}

For the mechanistic ODE models, we used the training strategies described above, but replaced Adam with the \texttt{Fides} Newton trust-region optimizer because \texttt{Fides} has been shown to perform well for mechanistic ODEs~\citep{frohlich_fides_2022}. As Fides has adaptive step-length, no learning-rate tuning was required. One tuning parameter required by \texttt{Fides} is the Hessian approximation. Here, we used a Gauss--Newton Hessian approximation, which has been shown to outperform BFGS approximations for the mechanistic ODE models considered here~\citep{persson_petabjl_2025}.

\subsection{Hyperparameter tuning and selection}
\label{sec:tuning}

We report the hyperparameter ranges explored for each method together with the final selected values, including those used for the mechanistic models.

\subsubsection{UDE and NODE benchmarks}

For the UDE/NODE benchmark models (Lotka--Volterra UDE, Lotka--Volterra NODE, SIR UDE, and pendulum UDE), hyperparameters for each training strategy were tuned using a two-step linear sweep. We first tuned the learning rate, as poor learning-rate choices led to poor performance regardless of the other hyperparameters. After fixing the learning rate, we swept the remaining hyperparameters around central values and selected one value at a time. Configurations were compared using the training/interpolation loss across $100$ random multistarts, summarized by the median. We generally selected the value with the lowest interpolation median loss, except for output regularization (Eq.~\ref{eq:output_reg}). Because output regularization is intended to prevent the neural network from dominating the dynamics and absorbing mechanistic components, we selected the largest $\lambda_O$ whose training loss was not significantly worse than the lowest-median configuration according to a Wilcoxon rank-sum test. We did not perform a full grid search due to the prohibitive computational cost. The explored values and selected settings are reported in Tabs.~\ref{tab:adam_tuning}--\ref{tab:cl_ms_tuning}.

Notably, output regularization (Eq.~\ref{eq:output_reg}) was generally tuned to be large only for the Lotka--Volterra UDE (Tabs.~\ref{tab:adam_tuning}--\ref{tab:cl_ms_tuning}). For the Lotka--Volterra NODE and pendulum models without mechanistic parameters, the neural network must represent the full dynamics, so penalizing its output may hinder learning. In the SIR UDE, the neural network is structurally restricted to a specific quarantine-related component (Eq.~\ref{eq:sir_eq}), making it less likely to replace the full mechanistic dynamics. In the SIR UDE, the neural network is structurally restricted to a specific quarantine-related component (Eq.~\ref{eq:sir_eq}); strong output regularization may therefore suppress the mechanism it is intended to learn, rather than prevent it from replacing the full dynamics. In contrast, for the Lotka--Volterra UDE, the neural network can in principle explain most of the right-hand side, for example by driving the mechanistic parameters toward zero.

\begin{table}[ht]
    \centering
    \caption{Adam training strategy hyperparameter tuning. Shown are the linear search spaces and selected values used for the Adam strategy. Tuned hyperparameters are the learning rate $\eta$ and the output regularization strength $\lambda_O$ (Eq.~\ref{eq:output_reg}).}
    \label{tab:adam_tuning}
    \begin{tabularx}{\linewidth}{llXc}
    \toprule
    Model & Parameter & Search space & Selected value \\
    \midrule
    \multirow{3}{*}{LV-UDE} & $\eta$ & [$0.001$, $0.01$, $0.05$, $0.1$] & $0.05$ \\
    & \multirow{2}{*}{$\lambda_O$} & [$0.0$, $0.1$, $1.0$, $10.0$, $100.0$, $500.0$, $1000.0$, $3000.0$, $5000.0$, $10\,000.0$, $30\,000$, $50\,000.0$] & \multirow{2}{*}{$1000.0$} \\
    \midrule
    \multirow{3}{*}{LV-NODE}
    & $\eta$ &
    [$0.001$, $0.005$, $0.01$, $0.05$, $0.1$] &
    $0.01$ \\
    & \multirow{2}{*}{$\lambda_O$}  &
    [$0.0$, $0.1$, $1.0$, $10.0$, $100.0$, $500.0$, $1000.0$, $5000.0$, $10\,000.0$, $50\,000.0$] &
    \multirow{2}{*}{$0.0$}   \\
    \midrule
    \multirow{2}{*}{SIR UDE} & $\eta$ & [$0.001$, $0.005$, $0.01$, $0.05$, $0.1$] & $0.005$ \\
    & \multirow{1}{*}{$\lambda_O$}  &
    [$0.0$, $0.0001$, $0.001$, $0.01$, $0.1$, $1.0$] &
    \multirow{1}{*}{$0.001$}   \\
    \midrule
    \multirow{2}{*}{Double pendulum} & $\eta$ & [$0.001$, $0.005$, $0.01$, $0.02$ $0.05$, $0.1$] & $0.02$ \\
    & \multirow{1}{*}{$\lambda_O$}  &
    [$0.0$, $1.0$, $500.0$, $500.0$] &
    \multirow{1}{*}{$0.0$}   \\
    \bottomrule
\end{tabularx}
\end{table}

\begin{table}[ht]
    \centering
    \caption{Curriculum learning (CL) training strategy hyperparameter tuning. Shown are the linear search spaces and selected values used for the CL strategy. Tuned hyperparameters are the learning rate $\eta$, number of curriculum stages $n_c$ and the output regularization strength $\lambda_O$ (Eq.~\ref{eq:output_reg}).}
    \label{tab:cl_tuning}
    \begin{tabularx}{\linewidth}{llXc}
    \toprule
    Model & Parameter & Search space & Selected value \\
    \midrule
    \multirow{4}{*}{LV-UDE} & $\eta$ & [$0.001$, $0.01$, $0.05$, $0.1$] & $0.01$ \\
    & $n_c$ & [$3$, $5$, $9$, $13$, $16$, $21$] & $13$ \\
    & \multirow{2}{*}{$\lambda_O$} & [$0.0$, $0.1$, $1.0$, $10.0$, $100.0$, $500.0$, $1000.0$, $3000.0$, $5000.0$, $10\,000.0$, $30\,000$, $50\,000.0$] & \multirow{2}{*}{$1000.0$} \\
    \midrule
    \multirow{4}{*}{LV-NODE}
    & $\eta$ &
    [$0.001$, $0.005$, $0.01$, $0.05$, $0.1$] &
    $0.01$ \\
    & $n_c$ & [$3$, $5$, $9$, $13$, $16$, $21$] & $21$ \\
    & \multirow{2}{*}{$\lambda_O$}  &
    [$0.0$, $0.1$, $1.0$, $10.0$, $100.0$, $500.0$, $1000.0$, $5000.0$, $10\,000.0$, $50\,000.0$] &
    \multirow{2}{*}{$0.0$}   \\
    \midrule
    \multirow{3}{*}{SIR UDE} & $\eta$ & [$0.001$, $0.005$, $0.01$, $0.05$, $0.1$] & $0.005$ \\
    & $n_c$ & [$3$, $5$, $9$, $13$, $16$, $21$] & $21$ \\
    & \multirow{1}{*}{$\lambda_O$}  &
    [$0.0$, $0.0001$, $0.001$, $0.01$, $0.1$, $1.0$] &
    \multirow{1}{*}{$0.0$}   \\
    \midrule
    \multirow{3}{*}{Double pendulum} & $\eta$ & [$0.001$, $0.005$, $0.01$, $0.02$ $0.05$, $0.1$] & $0.005$ \\
    & $n_c$ & [$3$, $5$, $9$, $13$, $16$, $21$] & $16$ \\
    & \multirow{1}{*}{$\lambda_O$}  &
    [$0.0$, $1.0$, $500.0$, $500.0$] &
    \multirow{1}{*}{$0.0$}   \\
    \bottomrule
\end{tabularx}
\end{table}

\clearpage

\begin{table}[ht]
    \centering
    \caption{Multiple shooting (MS) training strategy hyperparameter tuning. Shown are the linear search spaces and selected values used for the MS strategy. Tuned hyperparameters are the learning rate $\eta$, number of multiple shooting windows ($n_w$), continuity penalty ($\lambda_{\mathrm{cont}}$) and the output regularization strength $\lambda_O$ (Eq.~\ref{eq:output_reg}).}
    \label{tab:ms_tuning}
    \begin{tabularx}{\linewidth}{llXc}
    \toprule
    Model & Parameter & Search space & Selected value \\
    \midrule
    \multirow{6}{*}{LV-UDE} & $\eta$ & [$0.001$, $0.01$, $0.05$, $0.1$] & $0.05$ \\
    & $n_w$ & [$3$, $4$, $5$, $9$, $13$, $16$, $21$] & $16$ \\
    & \multirow{2}{*}{$\lambda_{\mathrm{cont}}$} & [$0.0$, $0.001$, $0.01$, $0.1$, $1.0$, $10.0$, $100.0$, $1000.0$] & \multirow{2}{*}{$10.0$}\\
    & \multirow{2}{*}{$\lambda_O$} & [$0.0$, $0.1$, $1.0$, $10.0$, $100.0$, $500.0$, $1000.0$, $3000.0$, $5000.0$, $10\,000.0$, $30\,000$, $50\,000.0$] & \multirow{2}{*}{$1000.0$} \\
    \midrule
    \multirow{6}{*}{LV-NODE}
    & $\eta$ &
    [$0.001$, $0.005$, $0.01$, $0.05$, $0.1$] &
    $0.05$ \\
    & $n_w$ & [$3$, $4$, $5$, $9$, $13$, $16$, $21$] & $16$ \\
    & \multirow{2}{*}{$\lambda_{\mathrm{cont}}$} & [$0.0$, $0.001$, $0.01$, $0.1$, $1.0$, $10.0$, $100.0$, $1000.0$] & \multirow{2}{*}{$10.0$}\\
    & \multirow{2}{*}{$\lambda_O$} &
    [$0.0$, $0.1$, $1.0$, $10.0$, $100.0$, $500.0$, $1000.0$, $5000.0$, $10\,000.0$, $50\,000.0$] &
    \multirow{2}{*}{$0.0$}   \\
    \midrule
    \multirow{4}{*}{SIR UDE} & $\eta$ & [$0.001$, $0.005$, $0.01$, $0.05$, $0.1$] & $0.1$ \\
    & $n_w$ & [$4$, $5$, $6$, $9$, $16$] & $4$ \\
    & \multirow{1}{*}{$\lambda_{\mathrm{cont}}$} & [$0.0$, $0.1$, $1.0$, $10.0$, $100.0$, $1000.0$, $10\,000.0$] & \multirow{1}{*}{$10\,000.0$} \\
    & $\lambda_O$ &  [$0.0$, $0.0001$, $0.001$, $0.01$, $0.1$, $1.0$] &
    \multirow{1}{*}{$1.0$}   \\
    \midrule
    \multirow{4}{*}{Double pendulum} & $\eta$ & [$0.001$, $0.005$, $0.01$, $0.02$ $0.05$, $0.1$] & $0.05$ \\
    & $n_w$ & [$6$, $8$, $13$, $27$, $32$, $40$, $54$] & $13$ \\
    & \multirow{1}{*}{$\lambda_{\mathrm{cont}}$} & [$0.0$, $0.1$, $100.0$, $500.0$] & \multirow{1}{*}{$100.0$} \\
    & \multirow{1}{*}{$\lambda_O$}  &
    [$0.0$, $1.0$, $100.0$, $500.0$] &
    \multirow{1}{*}{$0.0$}   \\
    \bottomrule
\end{tabularx}
\end{table}

\begin{table}[H]
    \centering
    \caption{Curriculum multiple shooting (CMS) training strategy hyperparameter tuning. Shown are the linear search spaces and selected values used for the MS strategy. Tuned hyperparameters are the learning rate $\eta$, number of initial multiple shooting windows stages $n_w$, continuity penalty $\lambda_{\mathrm{cont}}$ and the output regularization strength $\lambda_O$ (Eq.~\ref{eq:output_reg}).}
    \label{tab:cl_ms_tuning}
    \begin{tabularx}{\linewidth}{llXc}
    \toprule
    Model & Parameter & Search space & Selected value \\
    \midrule
    \multirow{6}{*}{LV-UDE} & $\eta$ & [$0.001$, $0.01$, $0.05$, $0.1$] & $0.01$ \\
    & $n_w$ & [$3$, $4$, $5$, $9$, $13$, $16$, $21$] & $16$ \\
    & \multirow{2}{*}{$\lambda_{\mathrm{cont}}$} & [$0.0$, $0.001$, $0.01$, $0.1$, $1.0$, $10.0$, $100.0$, $1000.0$] & \multirow{2}{*}{$1.0$}\\
    & \multirow{2}{*}{$\lambda_O$} & [$0.0$, $0.1$, $1.0$, $10.0$, $100.0$, $500.0$, $1000.0$, $3000.0$, $5000.0$, $10\,000.0$, $30\,000$, $50\,000.0$] & \multirow{2}{*}{$5000.0$} \\
    \midrule
    \multirow{6}{*}{LV-NODE}
    & $\eta$ &
    [$0.001$, $0.005$, $0.01$, $0.05$, $0.1$] &
    $0.01$ \\
    & $n_w$ & [$3$, $4$, $5$, $9$, $13$, $16$, $21$] & $16$  \\
    & \multirow{2}{*}{$\lambda_{\mathrm{cont}}$} & [$0.0$, $0.001$, $0.01$, $0.1$, $1.0$, $10.0$, $100.0$, $1000.0$] & \multirow{2}{*}{$1.0$}\\
    & \multirow{2}{*}{$\lambda_O$} &
    [$0.0$, $0.1$, $1.0$, $10.0$, $100.0$, $500.0$, $1000.0$, $5000.0$, $10\,000.0$, $50\,000.0$] &
    \multirow{2}{*}{$0.0$}   \\
    \midrule
    \multirow{4}{*}{SIR UDE} & $\eta$ & [$0.001$, $0.005$, $0.01$, $0.05$, $0.1$] & $0.005$ \\
    & $n_w$ & [$4$, $5$, $6$, $9$, $16$]  & $16$ \\
    & \multirow{1}{*}{$\lambda_{\mathrm{cont}}$} & [$0.0$, $0.1$, $1.0$, $10.0$, $100.0$, $1000.0$] & \multirow{1}{*}{$0.0$} \\
    & $\lambda_O$ &  [$0.0$, $0.0001$, $0.001$, $0.01$, $0.1$, $1.0$] &
    \multirow{1}{*}{$0.0$}   \\
    \midrule
    \multirow{4}{*}{Double pendulum} & $\eta$ & [$0.001$, $0.005$, $0.01$, $0.02$ $0.05$, $0.1$] & $0.05$ \\
    & $n_w$ & [$6$, $8$, $13$, $27$, $32$, $40$, $54$] & $8$ \\
    & \multirow{1}{*}{$\lambda_{\mathrm{cont}}$} & [$0.0$, $0.1$, $100.0$, $500.0$] & \multirow{1}{*}{$0.1$} \\
    & \multirow{1}{*}{$\lambda_O$}  &
    [$0.0$, $1.0$, $100.0$, $500.0$] &
    \multirow{1}{*}{$0.0$}   \\
    \bottomrule
\end{tabularx}
\end{table}

\subsubsection{Mechanistic ODE benchmarks}

As described in Sec.~\ref{sec:experiments}, the mechanistic ODE models were primarily included to assess the robustness of the training strategies and their performance on mechanistic models. For these benchmarks, we therefore performed a small grid search over hyperparameter values and report results for the best-performing configuration, measured by the number of runs converging to the global optimum across 1000 multistarts (Fig.~\ref{fig:mech_benchmarks}). For curriculum learning (CL), we tested 2, 3, 4, and 5 curriculum stages ($n_c$). For multiple shooting (MS) and curriculum multiple shooting (CMS), we tested 2, 3, 4, and 5 initial windows ($n_w$), together with continuity-penalty values of 0.0, 1.0, 10.0, and 100.0 for $\lambda_{\mathrm{cont}}$. The only exception was the hard-to-train Elowitz model, for which we tested 2--9 windows and curriculum stages, and the Fujita and Zheng models, for which we tested only 2, 3, and 4 windows and curriculum stages because these models contain few measurements at unique time points. The best-performing values, used for Fig.~\ref{fig:mech_benchmarks}, are reported in Tab.~\ref{tab:mech_tuning}.

\begin{table}[ht]
    \centering
    \caption{Hyperparameters selected for each training strategy for the mechanistic ODE models. For CL, $n_c$ denotes the number of curriculum stages. For MS and CMS, $n_w$ denotes the number of initial multiple-shooting windows and $\lambda_\mathrm{cont}$ the continuity penalty weight.}
    \label{tab:mech_tuning}
    \begin{tabularx}{0.52\linewidth}{lcccccc}
    \toprule
    & \multicolumn{1}{c}{CL} & & \multicolumn{2}{c}{MS} & \multicolumn{2}{c}{CMS} \\
    \cmidrule(lr){2-2}
    \cmidrule(lr){4-5}
    \cmidrule(lr){6-7}
    Model
    & $n_c$
    &
    & $n_w$
    & $\lambda_\mathrm{cont}$
    & $n_w$
    & $\lambda_\mathrm{cont}$ \\
    \midrule
    Boehm     & 5 & & 5 & 0.1 & 5 & 0.1 \\
    Bruno     & 5 & & 4 & 100 & 4 & 100 \\
    Elowitz   & 6 & & 3 & 1 & 6 & 0 \\
    Fiedler   & 5 & & 4 & 100 & 5 & 10 \\
    Fujita    & 5 & & 2 & 0.1 & 2 & 0 \\
    Okuonghae & 2 & & 2 & 0.1 & 2 & 0.1 \\
    Schwen    & 2 & & 5 & 100 & 5 & 0.1 \\
    Sneyd     & 5 & & 4 & 100 & 5 & 100 \\
    \bottomrule
    \end{tabularx}
\end{table}

\clearpage

\end{document}